\documentclass[5p,authoryear]{elsarticle}
\usepackage[colorlinks,citecolor=red,urlcolor=blue,bookmarks=false,hypertexnames=true]{hyperref}
\usepackage{url}
\usepackage{comment}
\usepackage{ifthen}
\usepackage{amssymb}
\usepackage{balance}
\usepackage[normalem]{ulem} % for \sout
\usepackage[table]{xcolor}
\usepackage{array}
\usepackage{booktabs}
\usepackage{graphicx}
\usepackage{xspace}

\newenvironment{widequotation}{\list{}{\listparindent 1.5em \itemindent\listparindent
		\rightmargin 0pt \parsep 0pt plus 1pt}\item\relax}
{\endlist}

\def\signed#1{{\leavevmode\unskip\nobreak\hfil\penalty50\hskip2em
		\hbox{}\nobreak\hfil\raise-3pt\hbox{#1}%
		\parfillskip=0pt \finalhyphendemerits=0 \endgraf}}
\newsavebox\mybox
\newenvironment{aquote}[1]
{\savebox\mybox{(#1)}\begin{widequotation}\itshape``\ignorespaces}
	{\unskip''\signed{\usebox\mybox}\end{widequotation}}

\usepackage{paralist}
\newcounter{suggestion}
\newenvironment{suggestion}{\refstepcounter{suggestion}
	\vspace{2pt}
	\begin{compactitem}
		\item[(S\thesuggestion)] \it }
	{ \end{compactitem}\smallskip}

\newcommand{\aLalal}{a tool and process analyst\xspace}
\newcommand{\daba}{a technical expert\xspace}
\newcommand{\mjewrou}{a solution architect\xspace}
\newcommand{\sdfe}{a concept leader\xspace}
\newcommand{\ssdf}{a product owner\xspace}
\newcommand{\ssdfwer}{a solution architect\xspace}
\newcommand{\werer}{a tools architect\xspace}
\newcommand{\ererwe}{a functional architect\xspace}
\newcommand{\ererwedf}{a requirements manager\xspace}
\newcommand{\trtret}{a methods and tools expert\xspace}
\newcommand{\trtretsewrt}{A tool and process analyst\xspace}
\newcommand{\tdf}{A technical expert\xspace}
\newcommand{\tdfd}{A solution architect\xspace}
\newcommand{\tdfdd}{A tools and methods specialist\xspace}
\newcommand{\tdfddd}{A concept leader\xspace}
\newcommand{\tefd}{A product owner\xspace}
\newcommand{\tefdd}{A solution architect\xspace}
\newcommand{\tefdde}{A tools architect\xspace}
\newcommand{\tefdded}{A functional architect\xspace}
\newcommand{\tefddedd}{A requirements manager\xspace}
\newcommand{\tefddeddd}{A methods and tools expert\xspace}

\renewcommand\cite{\citep}
\usepackage{tikz}
\makeatletter
\def\@author#1{\g@addto@macro\elsauthors{\normalsize%
		\def\baselinestretch{1}%
		\upshape\authorsep#1\unskip\textsuperscript{%
			\ifx\@fnmark\@empty\else\unskip\sep\@fnmark\let\sep=,\fi
			\ifx\@corref\@empty\else\unskip\sep\@corref\let\sep=,\fi
		}%
		\def\authorsep{\unskip,\space}%
		\global\let\@fnmark\@empty
		\global\let\@corref\@empty  %% Added
		\global\let\sep\@empty}%
	\@eadauthor={#1}
}
\makeatother

\journal{Journal of Systems and Software}
\begin{document}
\title{Why and How to Balance Alignment and Diversity of Requirements Engineering Practices in 
Automotive}

\author{Rebekka Wohlrab\corref{cor1}\fnref{label2,sys}}
\cortext[cor1]{Corresponding author}
\ead{wohlrab@chalmers.se}
\author{Eric Knauss\fnref{label2}}
\ead{eric.knauss@gu.se}
\author{Patrizio Pelliccione\fnref{label2,laq}}
\ead{patrizio.pelliccione@gu.se}
\fntext[label2]{Chalmers $|$ University of Gothenburg, Gothenburg, Sweden}
\fntext[sys]{Systemite AB, Gothenburg, Sweden}
\fntext[laq]{University of L'Aquila, L'Aquila, Italy}


\begin{abstract}
	\begin{tikzpicture}[overlay]
	\node[draw, fill=white, thick, align=center, style={inner sep = 4}] (b) at (10,7){\vspace{0.5em}This is a preprint of the following article, accepted to the Journal of Systems and Software:\\
		Rebekka Wohlrab, Eric Knauss, Patrizio Pelliccione,
		Why and How to Balance Alignment and Diversity of Requirements\\ Engineering Practices in Automotive,	Journal of Systems and Software, 2019, 110516, ISSN 0164-1212,\\ \url{https://doi.org/10.1016/j.jss.2019.110516}.};
	\end{tikzpicture}
In large-scale automotive companies, various requirements engineering (RE) practices are used across teams.
RE practices manifest in Requirements Information Models (RIM) that define what concepts and information should be captured for requirements.
Collaboration of practitioners from different parts of an organization is required to define a suitable RIM that balances support for diverse practices in individual teams with the alignment needed for a shared view and team support on system level.
There exists no guidance for this challenging task.
This paper presents a mixed methods study to examine the role of RIMs in balancing alignment and diversity of RE practices in four automotive companies.
Our analysis is based on data from systems engineering tools, 11 semi-structured interviews, and a survey to validate findings and suggestions.
We found that balancing alignment and diversity of RE practices is important to consider when defining RIMs.
We further investigated enablers for this balance and actions that practitioners take to achieve it.
From these factors, we derived and evaluated recommendations for managing RIMs in practice that take into account the lifecycle of requirements and allow for diverse practices across sub-disciplines in early development, while enforcing alignment of requirements that are close to release.
\end{abstract}

\maketitle
% !TEX root =  main.tex
\section{Introduction} \label{sec:Introduction}
Scale has become an important research hotspot in requirements engineering, as the systems' size and complexity increase, {and} requirements originate from an increasing number of stakeholders and disciplines and need to be combined into a ``single coherent story''~\cite{Cheng2009}.
However, while efforts exist to create common and company-wide requirements engineering methods~\cite{Weber2002}, the need to acknowledge diversity and tailor requirements engineering methods to specific contexts has been  acknowledged~\cite{Davis2013}.
In the automotive domain in particular, practitioners need to find a balance between the diversity and alignment of requirements engineering practices~\cite{Wohlrab2018REFSQ}.
Diversity and alignment can be observed based on how requirements-related knowledge is created, changed, and maintained in artifacts (e.g., models or documents) by several different groups in an organization.
A common {\em Requirements Information Model} (RIM) can help to ``develop a 
common view about requirements'' and to create tool support~\cite{John1999}.
Practitioners see a benefit in standardizing artifact models for requirements engineering, but also the need to tailor models to individual projects~\cite{Mendez2015}.

\newcommand{\rqWhy}{ What factors motivate the need to support alignment and diversity in RIMs in large-scale automotive companies?\xspace}
\newcommand{\rqEnablers}{How do	RIMs enable the balance of alignment and diversity of 
	RE practices in large-scale automotive companies?\xspace}

\newcommand{\rqActions}{What actions can be observed when large-scale automotive companies balance alignment and diversity using their RIMs?\xspace}

\newcommand{\rqSuggestions}{What are suggestions for managing 
	RIMs to balance alignment and diversity of RE 
	practices?\xspace}

To the best of our 
knowledge, there exists no study that sheds light on the 
underlying reasons to 
balance alignment and diversity of RE practices in automotive.
We focus on this aspect in our first research question.

\indent\textbf{RQ1}: \textit{\rqWhy}
\vspace{0.5em}

As mentioned before, RIMs can be used to standardize RE 
practices, but can also be tailored to individual 
projects~\cite{Mendez2015}.
Our second research question is concerned with how RIMs 
enable the balance of alignment and diversity in practice.

\indent\textbf{RQ2}:  \textit{\rqEnablers}
\vspace{0.5em}

As any artifact and model, RIMs have lifecycles and are evolved over time.
While related work has explored how RIMs can be created, there exists a knowledge gap with respect to how RIMs are changed to achieve a balance of alignment and diversity of RE practices.

\indent\textbf{RQ3}: \textit{\rqActions} 
\vspace{0.5em}

Finally, to give actionable guidance to practitioners, we focus on suggestions to manage RIMs to achieve a balance of alignment and diversity in large-scale automotive requirements engineering:

\indent\textbf{RQ4}: \textit{\rqSuggestions}
\vspace{0.5em}

\begin{figure} [b]
	\centering
	\includegraphics[width=.7\linewidth]{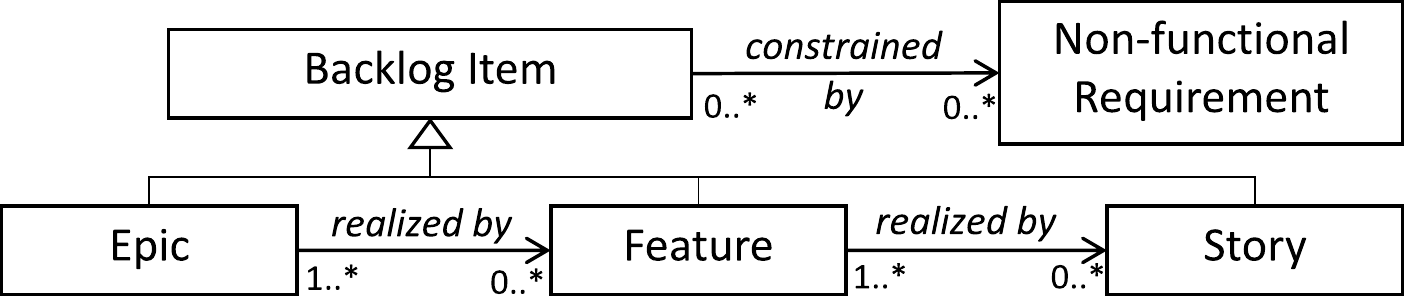}
	\caption{Excerpt of a requirements information model, adapted from~\cite{Leffingwell2011}}
	\label{fig:leffingwell}
\end{figure}

We performed a study using a mixed methods 
approach together with four automotive companies, including 
data analysis of a systems engineering tool, document 
analysis, 11 semi-structured interviews, and a survey with 19 
responses.
We contribute insights into how both alignment and diversity of RE practices are needed and supported by RIMs in practice.
Alignment is crucial in certain phases of the lifecycle of requirements, e.g., when the product is released.
Concrete requirements influence how much alignment and diversity is desired over time.
The RIM undergoes periods of change and stability, until elements of the RIM 
can become deprecated.
Our suggestions are to include key stakeholders, evaluate changes with few users, and focus on aligning high-level aspects.
We recommend to create new entity types only if special procedures, attributes, or relationships exist, support the creation of concrete requirements with minimal information, and favor training and flexibility over strong restrictions. Section~\ref{sec:Background} presents background information and Section~\ref{sec:RelatedWork} presents related work. In Section~\ref{sec:ResearchMethod}, we describe the research method. In Section~\ref{sec:why}--\ref{sec:suggestions}, we present our research findings. Section~\ref{sec:Discussion} concludes this paper with a discussion.

\section{Background} \label{sec:Background}
This paper relates to diversity and alignment in large-scale automotive RE
and information models in RE.

\subsection{Diversity vs Alignment in Automotive RE} 
\label{sec:RelatedWork:DiversityAlignment}
A requirements engineering practice is ``the use of a
principle, tool, notation, and/or method in order to perform any or all of the [...] activities'' related to eliciting, analyzing, documenting, verifying, and changing requirements~\cite{Davis2005}.
Diversity of requirements engineering practices refers to the heterogeneity of principles, tools, notations, and methods used in different groups in an organization.
Alignment refers to how similarly and consistently principles, tools, notations, and methods are used in different organizational groups.

In the automotive domain in particular, the heterogeneity of 
functions and quality attributes is a prevalent challenge for 
software and systems engineers~\cite{Pretschner2007}.
Especially in such a diverse domain, 
mechanisms are needed to consolidate requirements engineering 
practices of several teams and create sufficient alignment~\cite{Wohlrab2018REFSQ}.
Multiple technical domains are involved (e.g., 
entertainment or power train) that come with particular domain-specific 
issues~\cite{Weber2002}.
Thousands of engineers collaborate in large-scale distributed setups and need 
to fulfill a large variety of quality attributes (e.g., safety, performance, 
security, and usability)~\cite{Ebert2017}.
The identified challenges raise the need to create novel development approaches 
and tools that allow practitioners to develop cost-efficient products in a 
highly complex domain~\cite{Broy2007}.
The variety of disciplines and the lack of common interdisciplinary 
understanding was found to be a complicating issue in automotive 
RE~\cite{Liebel2018}.
A rather rigorous RE approach is needed to create high-quality products and 
support OEM-supplier relationships~\cite{Ebert2017}.

\subsection{Classification and Information Models in RE} 
\label{sec:RelatedWork:InfModels}
Humans like to categorize and classify things, as it allows them to create structures for their lives and work \cite{Bowker1999}.
The need to create a classification scheme \cite{ISO11179} for requirements that is both generic and adaptable has been identified more than 20 years ago \cite{Hochmuller1997}.
Several approaches to classifying or modeling requirements have been created since then~(e.g., \citet{Gorschek2006,Mendez2010}).
For instance, viewpoints can be used to classify requirements, considering  perspectives of different stakeholders~\cite{Finkelstein1992,Sommerville1997}.

In this paper, we consider Requirements Information Models (RIMs) as artifacts that 
describe (1) \textit{entity types} of information and concepts related to requirements 
engineering, (2) their \textit{relationships}, and (3) \textit{constraints} to 
create requirements-related knowledge.
Often, only one standardized model is used within a 
company~\cite{Mendez2011}, but with increased scale different organizational 
groups start to adapt the RIM or even to create a new one.
Figure~\ref{fig:leffingwell} shows an excerpt of a RIM~\cite{Leffingwell2011}.
It includes \textit{Backlog Item} as a main entity type that can be \emph{constrained by} \textit{Non-Functional Requirements}.
\textit{Epics}, \textit{Features}, and \textit{Stories} are more specialized entity types of \textit{Backlog Item}.
Other terms for RIM are requirements metamodel, reference model, or artifact model~\cite{Mendez2011,Mendez2010}.
A RIM for agile enterprises focuses on backlog items to organize teams' tasks~\cite{Leffingwell2011}.
Even though related work indicated that backlogs are ``informal models of work to be done'' rather than requirements specifications~\cite{Sedano2019}, a backlog does relate to requirements and this relationship should be covered by the RIM.
In this paper, we are interested in how a RIM is changed throughout its lifecycle.
As any artifact or software, RIMs have a lifecycle, i.e., a ``chain of activities, transformations, events, and artifacts to guide the full process'' that encapsulates all activities needed to ``conceive, develop, deploy, and maintain a software product''~\cite{Rodriguez2009}.

The concept of \emph{boundary objects} has recently been receiving increasing attention in software engineering 
(e.g.,~\citet{Sedano2019,Wohlrab2019JSME}).
Boundary objects establish a common understanding between groups without compromising each group's identity~\cite{Star1989} and can become apparent when social groups establish standards and categories and manifest them in information artifacts~\cite{Bowker1999}.
Examples of boundary objects include forms and standards~\cite{Star1989a}.
``Boundary objects arise directly from the problematics created when two or more differently naturalized classification systems collide''~\cite{Bowker1999}. % p 297
For instance, boundary objects emerge when residual categories emerge in a categorization:
as more and more stakeholders choose the ``Other'' category, the need to group these things in subgroups emerges and new categories arise as boundary objects.
Bowker and Star identified the need to understand how boundary objects are established and maintained, and what role classification schemes play as artifacts~\cite{Bowker1999}.
While boundary objects can be on the concrete artifact level, we focus on the meta level and how RIMs can be leveraged as boundary objects.
Moreover, this paper focuses on how concrete requirements adhere to RIMs and how both are changed in practice.

\section{Related Work} \label{sec:RelatedWork}
A broad spectrum of methods and practices exist for RE activities and representations of requirements~\cite{Laplante2017}.
The need to support diverse practices has been reported in globalized RE contexts with various tools~\cite{Bhat2006}, and in situations where both domain-specific and generalized solutions are needed~\cite{Cheng2009}.
Processes in requirements engineering cannot be standardized for all situations, but need to follow certain conventions~\cite{Serna2017}.
There exist different potential levels of rigor in RE:
no or heavy process, no or strict standards, no or heavy documentation, no or rigorous reviews.
Neither of the two extremes is right ``for all companies, or even for all projects within any one company''~\cite{Davis2013}.
RIMs are promising to look at when examining the trade-off between diversity and alignment of teams, as they have been used to standardize RE practices, but also to allow individual adjustments according to a project's needs~\cite{Mendez2015}.
Moreover, RIMs have been found useful to support communication between the members of multiple projects when discussing RE processes~\cite{Doerr2004}.
This study sheds light on how RIMs can be established and evolved over time and what the motivating factors of alignment and diversity are.

So far, the topic of requirements-related boundary objects is rather unexplored.
A field study~\cite{Hertzum2004} has examined the use of boundary objects in requirements engineering and their use to coordinate and align organizational groups.
Apart from this initial study, there exists limited empirical knowledge on requirements-related boundary objects and their use to achieve an alignment-diversity balance in large-scale RE practices.

In recent years, RE research has focused on agile development and continuous deployment (e.g.,~\cite{Niu2018, Schon2017}), that can facilitate collaboration and communication in large-scale agile development~\cite{Inayat2015}.
Kassab found that various RE practices are used for agile development contexts and that it is common to create and manage RE-related information in several tools (e.g., application lifecycle management tools)~\cite{Kassab2014}.
One of the conclusions of a systematic literature review in the area was that also in agile requirements engineering, a variety of artifact types are used, that heterogeneous agile RE approaches are common, and that better ways to create ``a shared understanding [...] among project members and stakeholders'' are needed~\cite{Schon2017}.
Our study contributes towards these goals as it analyzes how heterogeneous approaches can be supported, while creating a shared understanding across sites.
The need to centrally consolidate RE-related information in large-scale development and using a platform to make it accessible to a heterogeneous group of stakeholders has been identified~\cite{Fucci2018}.
This paper contributes to an understanding of the required level of diversity and alignment and can influence future development of tools and solutions for large-scale RE.

In automotive, several model-based solutions have been suggested to conduct requirements engineering~\cite{Boulanger2008, Pretschner2007, Braun2014}.
The automotive domain is similar to the domain of avionics systems engineering. Also in the domain of avionics systems engineering, an information model has been designed that allows requirements to be linked to justification, constraints, designs, and acceptance tests~\cite{Pearson1997}.
Weber and Weisbrod~\cite{Weber2002} described how an RE team in an automotive 
OEM introduced a company-wide modular RIM that allowed projects to adapt and 
tailor the model to their needs.
They identified and stressed the need to support diverse needs of teams with 
user-~and situation-specific views on requirements.
We analyze not only how RIMs can be introduced, but how they can be leveraged 
and evolved to balance alignment and diversity.

To be able to trace requirements, an upfront strategy is required that should ideally be tailored to individual projects~\cite{Rempel2013}.
In practice, however, needs change over time and upfront strategies should be evolved.
In this paper, we contribute to an understanding how RIMs, also including traceability-related information, changes over time and should be adapted throughout their lifecycles.

In recent years, the need to engage a growing number of people in requirements engineering activities has been identified, which led to the rise of the field of crowd-based requirements engineering~\cite{Groen2017}.
While our study focuses on different heterogeneous teams in an organization, instead of a diverse user base, the insights into alignment and diversity can also be useful when trying to establish boundary objects for crowd-based RE and consolidating the needs of different user groups.
% !TEX root =  main.tex
\section{Research Method} \label{sec:ResearchMethod}
We answer our research questions based on a \emph{mixed methods approach} \cite{Easterbrook2008}.
We follow a design in which qualitative data from interviews and quantitative data from the systems engineering tool are analyzed together.
The survey data was used as an additional source after the data from other sources had been analyzed, in a sequential design.

\subsection{Selected Participants}
We selected three automotive companies to shed light on the topic from different angles.
The automotive industry is chosen as the need for the balance of alignment and 
diversity is a particularly challenging issue in this domain.
As described in Section~\ref{sec:RelatedWork:DiversityAlignment}, a variety of 
disciplines are involved that come with particular issues and need to be 
consolidated to facilitate the creation of one integrated product.
Two companies (OEM1 and OEM2) are automotive Original Equipment 
Manufacturers (OEMs).
As the supplier-OEM relationship is a particular characteristic of the automotive domain, another company is an automotive supplier (SUP).
Parts of all companies use agile practices, with the SAFe framework being the 
most commonly used framework~\cite{Leffingwell2007}.
They all consist of at least 10 teams, which makes them 
``very large-scale'' 
development contexts~\cite{Dingsøyr2014a}.
Counting suppliers and all involved departments, hundreds of 
thousands stakeholders participate in the development process.
Moreover, we collaborated with a tool supplier (TOOL) developing a systems engineering tool used in automotive companies, and selected participants involved in the customization of RIMs at different customers.
The employees of the tool supplier are often main stakeholders when creating and managing RIMs and understanding the balance of alignment and diversity.
Table~\ref{tab:interviewees} shows characteristics of the interviewees.

\subsection{Systems Engineering Tool Data and Documentation}
We analyzed the data of a systems engineering tool from two of the automotive companies, focusing both on the RIM and concrete requirements.
The systems engineering tool allowed us to analyze, among other aspects, what entity types and relationships exist in the RIMs, and how often these are used to describe requirements.
We include descriptive statistics in this paper based on the data analysis.

To analyze further requirements-related documents, we leveraged data presented on the companies' intranets, powerpoint presentations related to RE processes, user guides, and manuals.
These documents are typically used for internal training purposes or to explain processes and methods.

Systems engineering tool data and documentation was most beneficial to answer RQ2 (how do RIMs enable the balance of alignment and diversity of RE practices).
Also the actions of balancing alignment and diversity (RQ3) can be partially observed using the data.
However, we did not do a longitudinal study in which we actually studied changes and actions in-depth over a longer period of time.
\subsection{Semi-structured Interviews}
Semi-structured interviews allowed us to collect rich qualitative data and explore connections between relevant factors influencing the topic under study.
In our companies, we selected key stakeholders that work with RIMs and/or initiatives to align RE practices in an organization.
There is only a limited number of potential interviewees who are knowledgeable and experienced in these areas.
Besides selecting representatives from all companies, we also interviewed specialists working at a tool supplier that supported the companies in configuring their RIMs.

To conduct the interviews, we created an interview 
guide\footnote{\url{https://rebrand.ly/intv_guide}}, 
including both open-~and 
closed-ended questions.
The interview questions include references to the related 
research questions.
The interviews' lengths were between 45 and 105 minutes, with an average of 62 minutes.
9 of 11 interviewees agreed with recording the interview and we created 
transcripts afterwards to allow for thorough data analysis.
For 2 of 11 interviews, we relied on detailed hand-written notes that we turned 
into a transcript immediately after the interviews with a fresh memory of what 
had been said.

To analyze the data, we carefully read through the transcripts several times to 
get familiar with the data.
Afterwards, we performed \emph{coding}, which is concerned with categorizing 
text chunks from an interview and labeling them with suitable 
terms~\cite{Creswell2008}.
We created a priori codes based on our research questions.
Examples of the a priori codes are ``\textit{Why Alignment}'', 
``\textit{Alignment Enablers}'', 
``\textit{Diversity Enablers}'', and ``\textit{Suggestions}.''

We used NVivo 12~\cite{NVivo2019} for the analysis, which allowed us to manage 
the large amount of collected data and search it more easily.
An \emph{editing approach} was used for the analysis~\cite{Runeson2009}.
We started with the initial set of a priori codes and created new codes, 
revised them, split, and merged codes.
77 codes were created in total, also capturing aspects that were not part of 
the initial a priori codes.
We describe the coding approach with the following example:
\begin{aquote}{\werer}
	There are some rules [to support alignment]. For the more complicated 
	rules, I want to skip them 
	during the typing phase and then delay it to a post processing work, after 
	everything is done.  
\end{aquote}
The statement deals with rules and consistency checks that support alignment in 
certain phases.
We connected this statement to the code ``Consistency checks'' under 
``Alignment Enablers'', as well as to 
``Dynamic levels of alignment during development'' under the top-level code 
``Suggestions.''

After having coded several interviews, we checked whether each code reflected 
only 
one central idea or whether new codes should be established~\cite{Tesch1990}.
During the analysis, we made sure that a \emph{chain of evidence} was 
established~\cite{Runeson2009}.
We organized a coding workshop to discuss the codes with their relations.
We printed the codes on paper, together with the number of 
interviews in which the code was applied.
This allowed us to discuss the meanings of all codes and understand how 
commonly used they were.
We grouped the sheets of paper on a table, discussed relations, and identified 
themes.
We created a story to report on the findings for each research question.
We provide summaries of our main findings in boxes in the respective sections.
\begin{table}
	\begin{footnotesize}
		\centering
		\caption{Interviewees with their companies and experience}
		\label{tab:interviewees}
		\begin{tabular}{>{\raggedright}p{0.015\linewidth}>{\raggedright}p{0.11\linewidth}>{}p{0.55\linewidth}>{\raggedright}p{0.13\linewidth}}
			\toprule
			& Company & Interviewee & Experience \tabularnewline \midrule
			1 & SUP & Tool and process analyst & 5 years \tabularnewline
			2 & SUP & Methods and tools expert & 31 years \tabularnewline
			3 & TOOL & Technical expert & 34 years \tabularnewline
			4 & TOOL & Solution architect & 23 years \tabularnewline
			5 & OEM1 & Tool and methods specialist & 34 years \tabularnewline
			6 & OEM1 & Requirements manager & 20 years \tabularnewline
			7 & OEM1 & Functional architect & 13 years \tabularnewline
			8 & OEM2 & Solution architect & 33 years \tabularnewline
			9 & OEM2 & Concept leader & 20 years \tabularnewline
			10 & OEM2 & Product owner for customized tool solution & 24 years \tabularnewline
			11 & OEM2 & Tools architect & 31 years \tabularnewline
			\bottomrule
		\end{tabular}
	\end{footnotesize}
\end{table}
\subsection{Survey}
Based on the findings of the interviews, we collected preliminary answers to all research questions.
We created a survey\footnote{\url{https://rebrand.ly/survey-que}} to validate the research findings (member checking) and gather additional data from other experts.
To strengthen the focus of our survey questions to our research goals, we did not include general questions on our participants' daily work, but rather included findings from the questions in the interview guide that were directly linked to research questions (14 of 17 questions).
For all questions, the relation to our RQs was indicated.
We did a pilot run with a tool expert, and sent it to all interviewees plus 13 additional experts in the area.
These experts worked at the companies, plus one additional automotive OEM and an automotive supplier.
They were suggested by other participants based on their experience with RIMs.
Of our respondents, 6 worked at automotive suppliers, 8 at OEMs, and 5 at tooling companies.
The respondents could choose multiple roles, 12 selected tools and methods expert, 5 selected architect, and 4 developer.
The survey included Likert-scale 
questions~\cite{Likert1932} and open-ended questions to be 
answered in a text 
field.
We used an online survey and received 19 out of 24 responses.
For the analysis, we used R~\cite{RProject2019}, as well as NVivo~\cite{NVivo2019} for the qualitative analysis of comments.
\subsection{Threats to Validity}
We discuss threats to validity for mixed methods research by presenting threats to the qualitative and quantitative methods used, as well as threats arising from the combination of the two~\cite{Wohlin2012,Ihantola2011}.

\textit{Construct validity}
is concerned with the appropriateness of our measurement tools for the topic being studied.
It is potentially compromised by different interpretations of terms and constructs that our study focused on.
An example of a term that could be interpreted differently was ``requirements engineering practice''.
We clarified the term by referring to the definition of requirements engineering practices that was also mentioned in Section~\ref{sec:Background}~\cite{Davis2005}.
Moreover, the term ``requirements information models'' might have been misunderstood by participants.
To mitigate misunderstandings in the interviews, we also mentioned alternative terms (e.g., metamodel, artifact model) and gave concrete examples to clarify the concept.
Also the survey started with an initial definition of RIMs using a concrete example.
We made sure that the definition and examples were in line with the RIMs used in the systems engineering tool under study, so that also the systems engineering tool data and documentation could be analyzed based on the same constructs.
The consistent use of constructs and theories in all used methods contributed to the overall construct validity of our research design.

\textit{Internal validity}
is concerned with confounding factors influencing the relationship between variables, treatment, and results obtained.
We do not aim to arrive at conclusions about the impact of a treatment on certain variables, but explore the topic more openly.
We focused on giving contextual information in the description of the participating companies and authentically reporting on the findings~\cite{Ihantola2011}.
Internal validity has been compromised, for instance, by incompletely identified aspects that motivate alignment and diversity.
To capture initially unconsidered factors, we used data triangulation, using data from various quantitative and qualitative sources.
Especially the interviews and use of systems engineering tool data allowed us to explore the topic without being restricted to variables from the start.
To improve internal validity, we explicitly asked for additional relevant factors in the interviews and survey.
Moreover, we aimed to reduce researcher bias by involving several researchers in the study and discussing findings throughout the process.

\textit{Conclusion validity} is concerned with wrong conclusions about relationships in our findings---either finding relationships that do not actually exist or missing relationships.
For instance, it might have happened that there exist unconsidered relationships between factors and the balance of alignment and diversity.
The variety of research methods in our mixed methods design helped to improve conclusion validity and allowed us to triangulate findings related to existing or missing relationships.
However, conclusion validity could be compromised by inaccurate measurement instruments.
For the questionnaire, potential threats to reliability are that questions might not have been presented in the right order or that the questionnaire took too long~\cite{Ihantola2011}, which might have influenced conclusion validity.
We started with a general explication of RIMs, followed by questions in the order of our research questions, and final demographic questions.
Throughout the questionnaire, 13 text fields were used for comments and further suggestions.
The questionnaire took 27 minutes on average, with a minimum of 9 minutes and a maximum of 60 minutes.
We expect that constructing a questionnaire with an understandable structure and tolerable length helped us arrive at correct conclusions about relationships related to the phenomenon under study.
When conducting systems engineering tool data analysis, we aimed to explore data in various ways (e.g., how often different entity types are instantiated) in parallel to conducting interviews, but could have missed relevant findings and relationships between factors.
For interviews, threats involve issues prohibiting us from accurately investigating relevant relationships (e.g., a too strict interview guide or a lack of asking subsequent questions to understand underlying objectives and motivations of interviewees).
To create an accurate interview guide, we created traceability between the interview questions and the research questions.
The interview transcripts helped us to conduct systematic analysis of data and trace findings to evidence from the data.

\textit{Reliability} is concerned with the consistency of our research method and whether researchers repeating the study would arrive at the same conclusions.
We aimed to improve reliability by aiming for transparency about our research method and deduction of findings by providing our instruments for data collection as separate documents.
We describe our analysis approach and use quotes for our findings to establish a clear chain of evidence.

\textit{External validity}
is concerned with generalizing research findings to other contexts.
In this mixed methods study, we involved a limited number of participants that all operate in their specific environments and points in time in the development processes.
There do not exist many practitioners who are knowledgeable in the area of RIMs and how to use them to balance the alignment and diversity of RE practices.
For the systems engineering tool data, we only analyzed data from two companies.
To mitigate the threat when performing semi-structured interviews, we collaborated with four companies and participants with different roles to consider several perspectives of the topic.
The survey allowed us to collect data from further companies.
We described the contexts of the companies so that other practitioners can compare characteristics and see what findings might be transferable.
In Section~\ref{sec:Discussion}, we discuss how transferable the findings might be to other contexts and domains.
% !TEX root = main.tex
\section{RQ1: Reasons for Alignment and Diversity} \label{sec:why}
This section answers RQ1: \emph{\rqWhy}

We found that when analyzing motivating factors for the alignment-diversity balance, there exist several factors supporting the importance of alignment, as well as of diversity.
Alignment and diversity can be combined and are not necessarily opposites.
We answer the following two sub-research questions, before summarizing our answers to RQ1 on a more general level:

\noindent\textbf{RQ1.1}: What factors motivate the need to support alignment in RIMs in large-scale automotive companies?

\noindent\textbf{RQ1.2}: What factors motivate the need to support diversity in RIMs in large-scale automotive companies?

We first describe our findings based on interviews, systems engineering tool data, and documentation.
In Section~\ref{sec:RQ1:reasons}, we summarize the findings and present the results of the survey regarding RQ1.
\subsection{RQ1.1: Motivating the Need for Alignment}
In the following, we present reasons for alignment in RIMs:
\begin{center}
	\begin{small}
		\begin{tabular}{|p{.9\columnwidth}|}
			\hline
			Alignment is mostly motivated by the need to facilitate integration, establish a common language, increase the quality of requirements, and adhere to standards. 
			\\ \hline
		\end{tabular}
	\end{small}
\end{center}\subsubsection{Facilitated Integration}
Six interviewees from all companies stated that an aligned RIM is needed to facilitate the integration of different functions and components, both internally and with suppliers.
Basing the specification of requirements on a common ground helps creating a recognizable structure to facilitate the integration of the work of different teams.
\tefdd stated that ``a big stakeholder is the continuous integration machine'' that requires users to follow the RIM and create prescriptive information understandable by a machine.
\tdfddd stressed that also when exchanging data with other tools and suppliers, it is crucial to align RIMs and facilitate the integration of the work products of different teams.
\trtretsewrt explained that this is especially useful when different components are created by different teams:
\begin{aquote}{\aLalal}
	We also have contracts on a requirements level, that both state what a component guarantees and what it requires.
\end{aquote}
If these contracts are captured in a formal model, they have to be adhered to in order to facilitate (continuous) integration.
\subsubsection{Common Language}
RIMs help with the coordination between teams, {mitigate} misunderstandings, and {support} the efficiency and effectiveness of an automotive company.
These aspects were explicitly mentioned by seven interviewees.
\begin{aquote}{\sdfe}
	The important thing is one information model, so that everyone speaks the same language. Electrical engineering, mechanical engineering, and so on. When you talk about requirements, then you know what you talk about.
\end{aquote}
\tefdded pointed out that ``it is necessary to have some common [elements], that we have a fairly common understanding of what they represent'', to make it easier to communicate between different teams and use common terms.
\subsubsection{Better Quality}
Five of 11 interviewees stressed that creating requirements that follow a common 
RIM increases their quality.
For this purpose, many large-scale companies support initiatives for common RE practices---not only with respect to the RIM, but, in general, by defining guidelines and styles.
\tefdde stated that ``standardized ways do not lead to quality on their own.''
Three interviewees from OEMs pointed out that especially the testability of requirements raises the need to establish common practices.
For instance, a RIM can support practitioners to create requirements on the right levels of granularity and establish consistent relationships to test cases.
Our interviewees considered it beneficial to follow these practices in a consistent way throughout the organization, which can be encouraged by RIMs as boundary objects.
Several interviewees also pointed out that better quality could be supported by tooling:
\begin{aquote}{\ererwedf}
	[Better tool features] would help, suggesting that you follow certain rules. [...] Sometimes we give a lot of freedom and the requirements are not so good.
\end{aquote}
\subsubsection{Standards} \label{sec:why:alignment:standards}
As safety is a prevalent concern in the automotive domain, common methods are followed to ensure compliance with ISO~26262~\cite{ISO26262}.
This point was explicitly stressed by three interviewees, both from the tool supplier, the supplier, and OEM1.
\tdf explained that it was necessary to ``formalize a lot of information.''
Dedicated parts of the RIM were created to support the analysis of hazardous operational situations and the derivation of safety goals and requirements.
\tdfdd described the need to follow strict processes for these parts and work in aligned ways throughout the company.
In a supplier company, safety documentation should also be aligned to ``communicate it to customers'' (\trtret).

\subsection{RQ1.2: Motivating the Need for Diversity}
Diversity is mostly motivated by the variety of disciplines involved in automotive engineering, the methods, natures of functions, and different techniques for elicitation.
This section presents reasons for diversity in RIMs:
\begin{center}
	\begin{small}
		\begin{tabular}{|p{.9\columnwidth}|}
			\hline
			Diversity is mostly motivated by the variety of disciplines involved in automotive engineering, the methods, natures of functions, and different methods for elicitation.
			\\ \hline
		\end{tabular}
	\end{small}
\end{center}
\subsubsection{Variety of Disciplines}Five of 11 interviewees named that different disciplines have different needs when it comes to RE practices.
\tefd elaborated on the differences between mechanical engineering and other disciplines and that the general RIM has to be adapted to fit to the needs of mechanical engineers: ``This way of thinking does not make sense for mechanical engineers.''
Also, \sdfe saw challenges with consolidating the needs of different disciplines and supporting different methods used in these fields.
\begin{aquote}{\sdfe}
	Interfaces can be electrical, digital, analog, ... It's not easy to do that when you work with people from all areas. Some have never seen hardware, some have never seen software.
\end{aquote}
The company decided to support diverse means of describing interfaces in the 
RIM, capturing the needs of all disciplines.
\tefd explained how interfaces in mechanical engineering describe, for 
instance, to how ``a seat is connected to the floor'' and are modeled in 
computer-aided design (CAD) models.
In electrical engineering, interfaces between ECUs (Electronic Control Units) 
are captured in signal databases, indicating what signals with data types and initial values are used.
OEM1 uses a dedicated change management system to keep track of 
requests to change signals in these databases and their requirements.
Interfaces in the software engineering domain can refer to abstract entity types 
defining method signatures that can be implemented by classes.
Using the RIM as a boundary object, practitioners can create artifacts of the entity type Interface, with a common understanding across sites, but also select precise subtypes of Interface to meet the specific interpretations of teams.
\subsubsection{Different Methods}
Three interviewees stated that RIMs should support both plan-driven and agile ways of working.
In fact, all of the participating companies are transitioning to agile methods.
Different methods are also used in different disciplines, as mentioned by five interviewees.
For instance, mechanical engineers at OEM2 create CAD models and describe 
product-related information 
in a specialized product lifecycle management (PLM) tool, whereas software 
development teams 
at SUP write source code in an integrated development environment, version the 
source code in the version control system 
Git\footnote{\url{https://git-scm.com/}}, and use the issue and project 
tracking software JIRA to keep track of 
changes\footnote{\url{https://www.atlassian.com/software/jira}}.
\tefd explained that, for instance, the start of production is less relevant for software developers working with continuous deployment than for other roles.

\tefdded elaborated that requirements are used for several purposes: In the traditional way of working in automotive, projects are concerned with evolving sets of requirements to address a defined purpose of a project.
In scenarios where suppliers and OEMs collaborate, the concrete methods change depending on the tools, individuals, functionality, legal contracts, and business relationships.
Furthermore, product documentation needs to be created and maintained, defining what requirements the product fulfills and serving as a reference for maintenance and aftermarket purposes.
When adopting agile methods, a backlog with epics, features, and stories is typically used to specify what software or systems aspects should be changed or added in a certain time interval.
A RIM should support these different ways of working with projects, products, and backlogs.

\subsubsection{Different Nature of Functions} \label{sec:sec:differentNature}
The different characteristics of functions are also reflected in the RIM and 
the entity types that should be specified.
This aspect came up in three interviews.
At OEM1, there exist functions for which the contexts in which a vehicle is 
situated is absolutely crucial for the requirements (e.g., the way headlamps 
should work depend on the road conditions, time of the day, weather, and 
location).
For other functions, these contexts are not important to specify, but user 
interface requirements should be described and modeled (e.g., for the central 
display).
Some functions require detailed descriptions of the scenarios, i.e., every step 
involved in the execution of a function.

At OEM2, use cases can consist of high-level descriptions of a function's purpose 
or formal description of a course of events:
\begin{aquote}{\werer}
	For a phone, use cases might be enough, but for control [systems engineering], you need scenarios to describe the expected behavior.
\end{aquote}

Moreover, it matters whether a function is a customer-specific or a generic 
function.
\trtretsewrt saw ``a difference between customer-specific and generic functions and 
how we find a way to implement the customer-specific functions.''
For customer-specific functions, the RIM should allow for the inclusion of 
particular details that facilitate the integration in the customer's end 
product.
A RIM needs to support ways of capturing contexts, user 
interface requirements, scenarios, or customer-specific details for some 
functions, but not necessarily in the same way for all functions.
\subsubsection{Creative Tasks and Elicitation}
When it comes to elicitation, different RE practices are supported, that also require diverse tool and modeling support.
Six interviewees gave examples of how use cases can be modeled with different styles: the more formal description with basic course of events, but also high-level use case summaries.
At OEM1, it is also possible to create state charts to describe high-level behavior as a part of a use case.
\tefddedd stated that ``tons of different methods'' for requirements elicitation have been described over the years.

\begin{figure}
	\centering
	\includegraphics[width=0.9\linewidth]{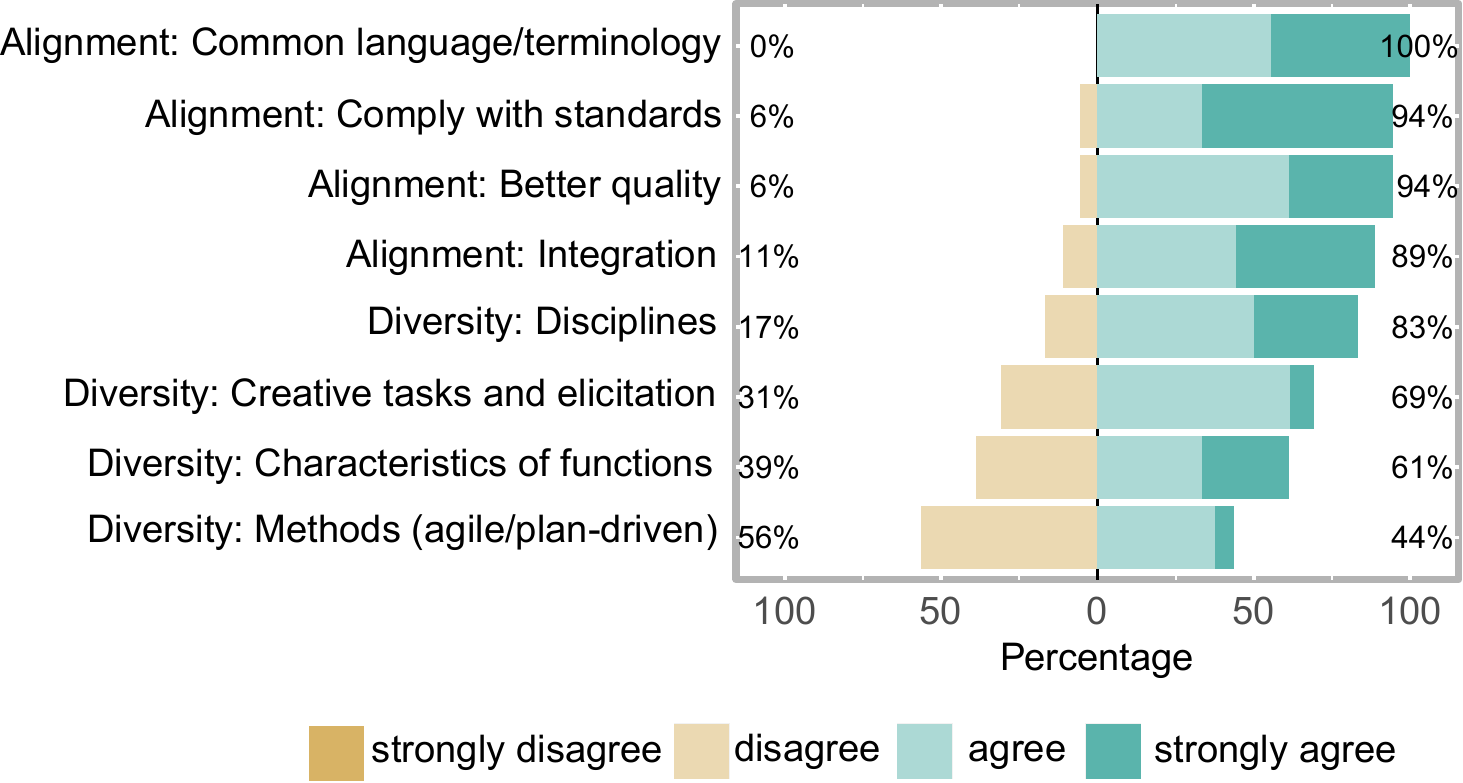}
	\caption{Survey responses w.r.t.~reasons for alignment and diversity: ``We need alignment/diversity in our RIM to support...'' (n = 19)}
	\label{fig:why}
\end{figure}

\subsection{Reasons for the Alignment-Diversity Balance} \label{sec:RQ1:reasons}
This section has presented factors supporting the need for alignment, as well as factors for diversity in RIMs.
We found that alignment is needed to facilitate integration, establish a common language, create requirements of better quality, and support the compliance with standards.
At the same time, different disciplines, methods, functions of different nature, and elicitation practices require diversity in RIMs and practices.
All factors are relevant and co-exist in large-scale automotive organizations.
\tefdd stated that finding the alignment-diversity balance ``is about finding the right, common [aligned parts of RIMs] and still some freedom, how to work within some boundaries.''

Figure~\ref{fig:why} shows the survey results regarding the need to support both alignment and diversity.
It can be seen that our respondents agree with the motivators for alignment, and gave more mixed answers on reasons for diversity.
A developer working at a supplier stated in the comments that ``\textit{diversity is good, but can also lead to too many styles and technical debt. Ideally the structure of different levels of requirements can be set early on, allowing not too much customizing.}''
We observed a difference in the roles: 75\% of the respondents who stated that they were tools and methods experts agreed or strongly agreed that different disciplines (e.g., mechanical engineering, software engineering) raise the need for diversity, whereas 100\% of the respondents that were no tools and methods experts agreed or strongly agreed with the statement.
Depending on the position in the company, diversity and alignment appear to be more or less observable.
20\% of the respondents working at the automotive supplier agreed with the statement that diversity is needed to support different development methods, whereas 56\% of the OEM employees agreed or strongly agreed with the statement.
As fewer disciplines are involved in the development at the supplier company SUP and more homogeneous development groups exist, the need to support different methods is not observed as much as in other types of companies.
However, supplier employees are more concerned with functions having different characteristics, e.g., generic functions vs.~customer-specific ones (see Section~\ref{sec:sec:differentNature}).
Respondents working at a supplier agreed or strongly agreed with the statement that diversity is needed because functions have different characteristics (67\%), whereas of the respondents employed at OEMs 44\% agreed or strongly agreed with the statement.

In the comments, also the need for aligning RE practices to support reuse was stressed.
A tools and methods expert from a supplier stated that alignment leads to a ``better dialogue and framework for the engineers to understand and get inspired by each other.''

% !TEX root =  main.tex
\section{RQ2: How to Enable Alignment and Diversity} \label{sec:enablers}
This section answers RQ2: \textit{\rqEnablers}

We understood that to enable the balance, mechanisms are needed to enable alignment, as well as diversity.
We answer the following sub-research questions in the following:

\noindent\textbf{RQ2.1}: How do RIMs enable alignment of RE practices in large-scale automotive organizations?

\noindent\textbf{RQ2.2}: How do RIMs enable diversity of RE practices in large-scale automotive organizations?

We leveraged systems engineering tool data and documentation to analyze how alignment and diversity are enabled.
To better understand rationales and motivations, we complemented this data with findings from interviews.
In Section~\ref{sec:survey:RQ2}, we summarize the findings and present the results of the survey regarding RQ2.
\subsection{RQ2.1: Enablers for Alignment}
With respect to aspects enabling alignment in RIMs, we arrived at the following finding:
\begin{center}
	\begin{small}
		\begin{tabular}{|p{.9\columnwidth}|}
			\hline
			RIMs support alignment by allowing to specify entity types and relationships, establishing common attributes, consistency checks, maturity levels, and Definition of Done criteria.\\ \hline
		\end{tabular}
	\end{small}
	
\end{center}

\begin{figure} [t]
	\centering
	\includegraphics[width=\linewidth]{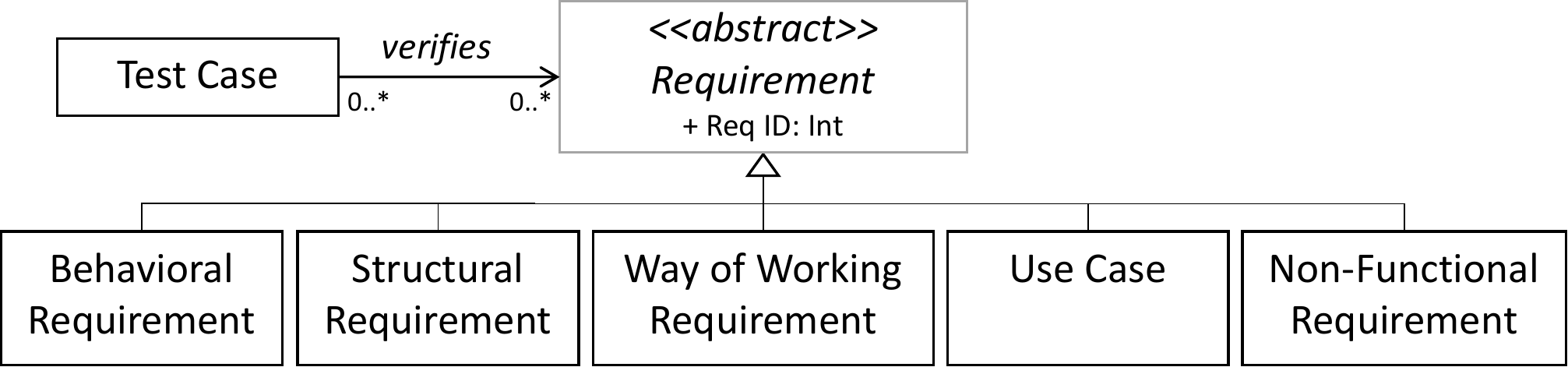}
	\caption{Excerpt of a minimal RIM at OEM2 with aligned aspects}
	\label{fig:alignedmm}
\end{figure}
\begin{figure}[b]
	\centering
	\includegraphics[width=\linewidth]{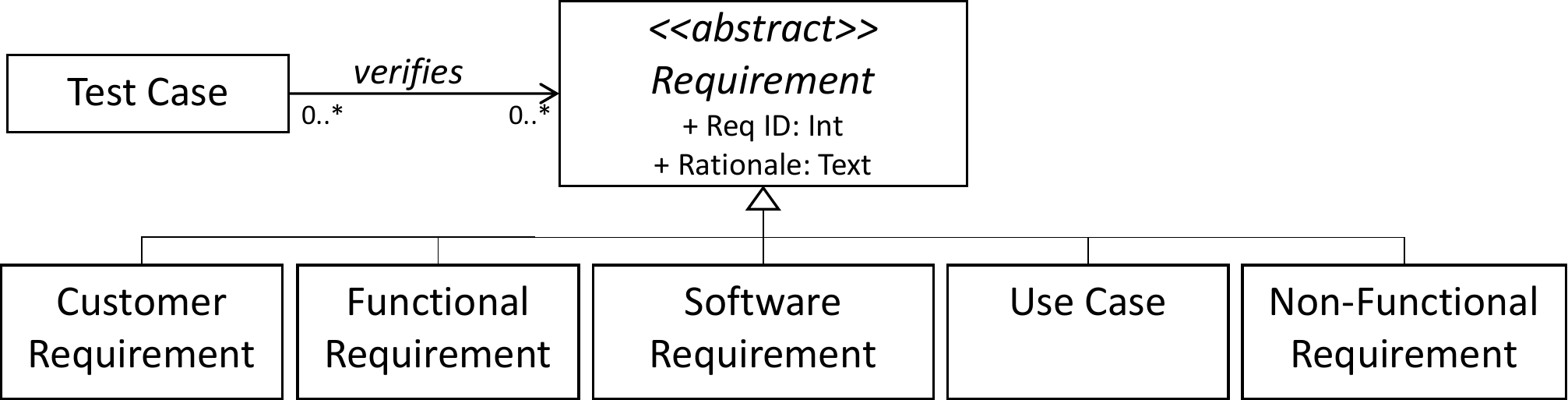}
	\caption{Excerpt of a minimal RIM at SUP with aligned aspects}
	\label{fig:sup}
\end{figure}

Figures~\ref{fig:alignedmm} and~\ref{fig:sup} show minimal excerpts of RIMs.
Each box represents an entity type in the model.
Attributes of an entity type are shown under the name of the entity type and are indicated with a plus sign (+).
There exist several subtypes of \textit{Requirement}.
Moreover, a relationship exists between \textit{Test Case} and \textit{Requirement}.

\subsubsection{Specification of Entity Types and Relationships}
Traditionally, automotive companies have worked based on documents with requirements specifications that were exchanged between different teams.
According to the experience of stakeholders of all companies, with the establishment of a common tool, aligned concepts need to be established and the semantics of pieces of information are described more precisely.
Seven interviewees mentioned this point explicitly and 89\% of the survey respondents agreed that alignment is supported by the specification of entity types and relationships.
\tdfd phrased this as follows:
\begin{aquote}{\mjewrou}
	[The systems engineering tool] is like a model of an organization. It is like a map of how things work.
	[...] In an organization based on documents, [...] you have more freedom and can just interpret things differently. But if you have this formal model with connections, and something is not connected, [...] then it needs to be fixed.
\end{aquote}
\emph{Specifying entity types in a RIM} enables the {alignment of RE practices}, because it establishes {common concepts with clear semantics to reflect the ways of working in the organization}.
Also the relationships are of crucial importance, to see how artifacts of different types are connected.
\emph{Specifying the relationships of entity types} has an {enabling} impact on {alignment of RE practices}, because {they establish ways to connect artifacts of different entity types based on their context and ensure traceability}.
According to \sdfe, a common, but limited set of requirements entity types helped to align how stakeholders view requirements and work with them.
Concretely, the common subtypes of OEM2 are \textit{Behavioral}, \textit{Structural}, \textit{Way of Working}, \textit{Non-Functional Requirement}, and \textit{Use Case}.
Behavioral requirements specify the behavior of a part of the vehicle, whereas structural requirements are concerned with the relation of different parts and their compositions.
Way of working requirements specify related working procedures that engineers should follow.

These requirement types are used in different phases in the development lifecycle, starting from early phases in which new functionality is described, to high-level design, and the concrete development of systems and components.
At OEM1 and SUP, entity types for each of these phases are defined, e.g., \emph{Functional Requirement}, \emph{Design Requirement}, or \emph{Software Requirement}.
At SUP, there exists a dedicated entity type for \emph{Customer Requirement}.
Table~\ref{tab:instanceSUP} gives an overview of the entity types from the example excerpt, the number of times they have been instantiated, and the number of relationships to instances of the entity types.
\begin{table}
	\centering
	\caption{Instance statistics of a minimal RIM at SUP}
	\label{tab:instanceSUP}
	\begin{footnotesize}
		\begin{tabular}{>{\raggedright}p{0.39\linewidth}>{}p{0.15\linewidth}>{\raggedright}p{0.32\linewidth}}
			\toprule
			Entity Type & No.~of\newline instances & No.~of relationships to instances \tabularnewline \midrule
			Test Case & 161,499 %nodes
			& 69,700 %defpart
			\tabularnewline 
			Customer Requirements & 2262 %nodes
			& 4911 %defpart
			\tabularnewline 
			Functional Requirements & 29,255 %nodes
			& 127,694 %defpart
			\tabularnewline 
			Software Requirements & 4995 %nodes
			& 12,682 %defpart
			\tabularnewline 
			Use Case & 7986 %nodes
			& 2316 %defpart
			\tabularnewline 
			Non-Functional Requirements & 5231 %nodes
			& 1613 %defpart
			\tabularnewline
			\bottomrule
		\end{tabular}
	\end{footnotesize}
\end{table}
In OEM2, the idea is that the level of abstraction or role for the development process can be understood from artifacts of other entity types pointing to the requirements (e.g., Function, System, or Component).

\subsubsection{Mandatory Attributes}
Requirement IDs and mandatory attributes, e.g., asking stakeholders to set the priority of a requirement, are ways to align RE practices.
Attributes should especially be standardized when the goal is to collaborate with suppliers or other companies.
Six interviewees mentioned that an identifier for a requirement is absolutely necessary for this purpose.
However, it was stressed as important that there should not be too many mandatory attributes.
\begin{aquote}{\ssdf}
	We don't want too many default compulsory attributes because people won't fill it in. And attributes should be self-explanatory. If it is compulsory you should get an error message.	
\end{aquote}
\subsubsection{Active Management through Consistency Checks and DoD Criteria}
Six interviewees stated that consistency checks, Definition of Done criteria, and maturity levels are mechanisms connected to the RIM that enable alignment.
They are most commonly used at SUP.
Consistency checks are especially used if the information is used in a prescriptive way, to create code or other artifacts.
\begin{aquote}{\mjewrou}
	If you compile it, it is more crucial. And then the awareness of the need to keep things connected and consistent becomes much stronger in the organization.
\end{aquote}

Another solution architect stressed that maturity levels could help ensuring 
consistency at an appropriate point in time:
\begin{aquote}{\ssdfwer}
	We could also add maturity levels and a workflow to check it. To reach status released, some condition should be fulfilled. [...] In an early stage, you can release items with a low maturity level, but then [...] a lot more checks will be done.
\end{aquote}

Also Definition of Done criteria enable alignment.
Typical criteria are to ensure that all entity types of the safety-related parts of the information model have been instantiated or that all software requirements have a relation to test cases.
\subsection{RQ2.2: Enablers for Diversity}
\begin{center}
	\begin{small}
		\begin{tabular}{|p{.9\columnwidth}|}
			\hline
			Diversity is enabled by supporting generic relationships, creating new subtypes with time, providing free text fields, and supporting several ways of organizing backlogs and projects.
			\\ \hline
		\end{tabular}
	\end{small}
	
\end{center}
\subsubsection{Generic Relationships}
Systems engineering tools typically support generic relationships between information of arbitrary entity types, either as a special relationship (``refers to'') or as hyperlinks.
Three interviewees explicitly reported that generic relationships are used to support diverse ways of modeling and following RE practices.
It is a powerful means to flexibly relate information.

According to \trtret and \ssdfwer, issues arise when circular references are created using these generic relationships or when information should be released.
The amount of control over information connected with relationships is limited.
\subsubsection{Creation of New Entity Types and Attributes}
In one of the used systems engineering tools, the metamodel can be extended at run-time, for instance, to add new entity types and attributes.
This feature was stressed by three interviewees.
\tefd explained that the entity type Interface got split into Mechanical, Electrical, and Software Interface to better capture diverse needs.
Also, both mandatory and optional attributes can be added and removed easily.
\tefdde from OEM2 suggested that the company should teach that mandatory attributes have to be filled in, but it should teach also how to add optional attributes whenever needed.
\subsubsection{Free Text Fields}
Four interviewees reported that descriptions in plain text enable diverse RE practices, as there do not exist any limitations with respect to the content, structure, or style of the texts.
Free text fields bring the advantage that people can add information in different ways, with varying levels of detail.
At the same time, it is not always followed as intended:
\begin{aquote}{\ssdfwer}
	People often express more than one requirement in one item. [...] We want single, clear requirements. There are also different use case styles.
\end{aquote}
\subsubsection{Flexible Use of Backlogs}
In six interviews, the flexible use of backlogs was mentioned as an enabler for diversity.
In all companies, the systems engineering tools under study are complemented by tools to manage backlogs, issues, and projects.
The scope of the aligned RIMs is limited to the systems engineering tools, while backlogs are typically managed in separate tools focusing on the ``delta'' (\ererwedf, OEM2).
While the information captured in the systems engineering tools should describe the product characteristics as a common reference, backlogs are rather used to organize what should be changed and for prioritization.
\begin{aquote}{\sdfe}
	The backlog describes what should be prioritized. And there the agile release trains should have the freedom. But [with the RIM] you have a well-defined interface that you should act towards.
\end{aquote}
\tefddeddd mentioned that in some cases, items exist in the backlog that point to the need of updating the requirements in the systems engineering tool.

\subsection{Enablers for the Alignment and Diversity} \label{sec:survey:RQ2}
In this section, we discussed several ways in which alignment and diversity are enabled.
Enablers for alignment are the formal specification of entity types and relationships in a RIM, mandatory attributes, consistency checks, and DoD criteria.
Diversity of RE practices is enabled by generic relationships, the extension of the RIM with new entity types and attributes, free text fields, and the flexible creation of backlog items.
The practices do not exclude each other and are used in different phases, as we will describe in the next section.

\begin{figure}
	\centering
	\includegraphics[width=\linewidth]{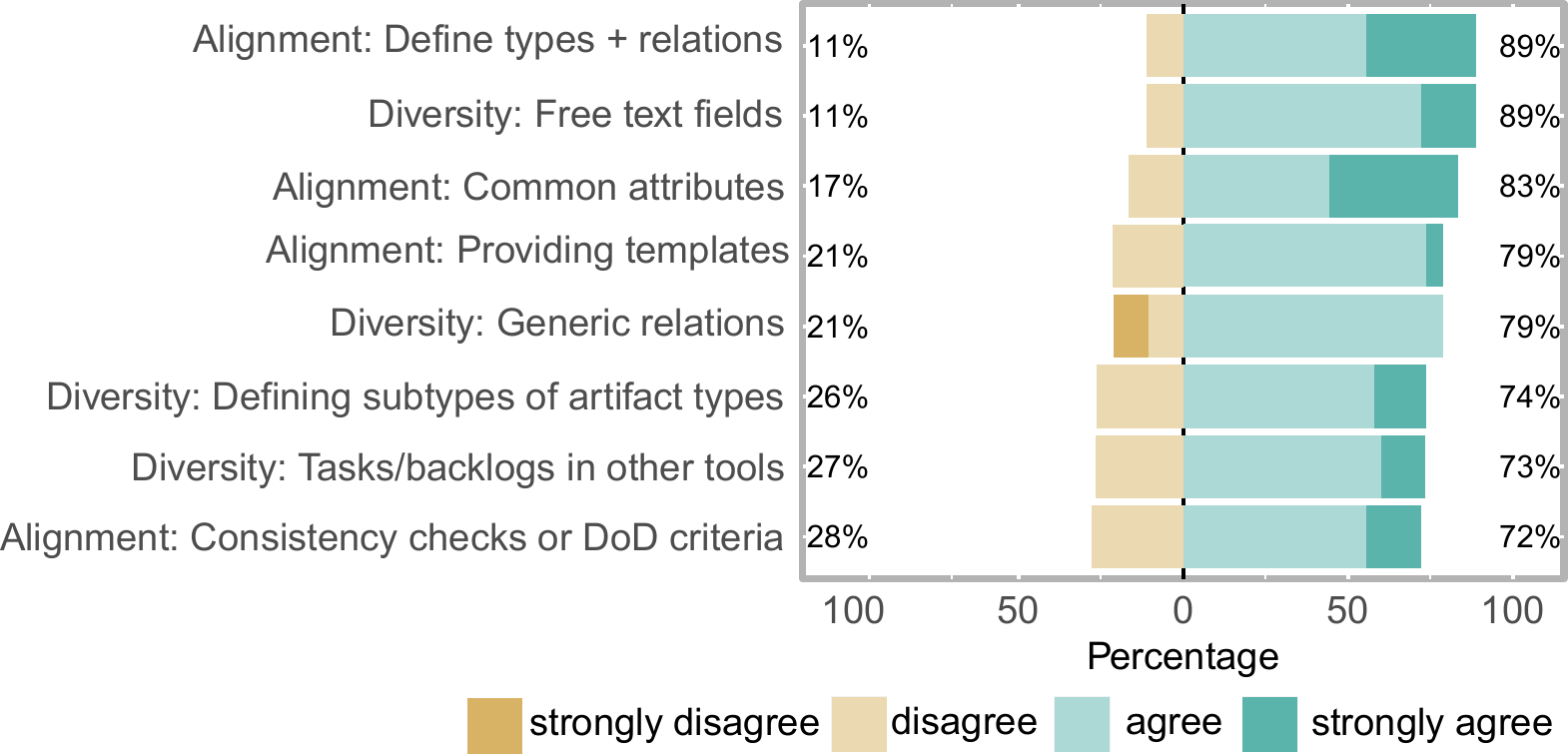}
	\caption{Survey responses about enabling alignment and diversity: ``We enable alignment/diversity using...'' (n = 19)}
	\label{fig:how}
\end{figure}

Figure~\ref{fig:how} shows the survey results regarding enablers for the alignment and diversity.
The respondents working at a supplier company gave slightly different answers than the remaining respondents: 100\% of the supplier employees stated that diversity is enabled by managing tasks and backlogs in other tools, by defining subtypes (100\% agreed or strongly agreed), and alignment is achieved by providing templates (100\% agreed or strongly agreed).
Of the respondents not employed at a supplier, 67\% agreed or strongly agreed with each of these statements.
The supplier company under study was the only one with a tool to manage backlogs which was used in all teams in the company (albeit with different tailored flavors in the concrete methods), and also templates were introduced with a company-wide strategy in this case.
For the other companies, tools and templates are introduced and recommended, but not with fixed company-wide rules.
These aspects influence what respondents regard as enablers for the alignment-diversity balance.
In the comments, the participants referred to how strongly they used the enablers so far.
A tools and methods experts from an OEM stated that ``we shall implement consistency checks but right now they are not in place.''

% !TEX root =  main.tex
\section{RQ3: Balancing Alignment and Diversity}  \label{sec:actions}
\begin{figure}
	\centering
	\includegraphics[width=\linewidth]{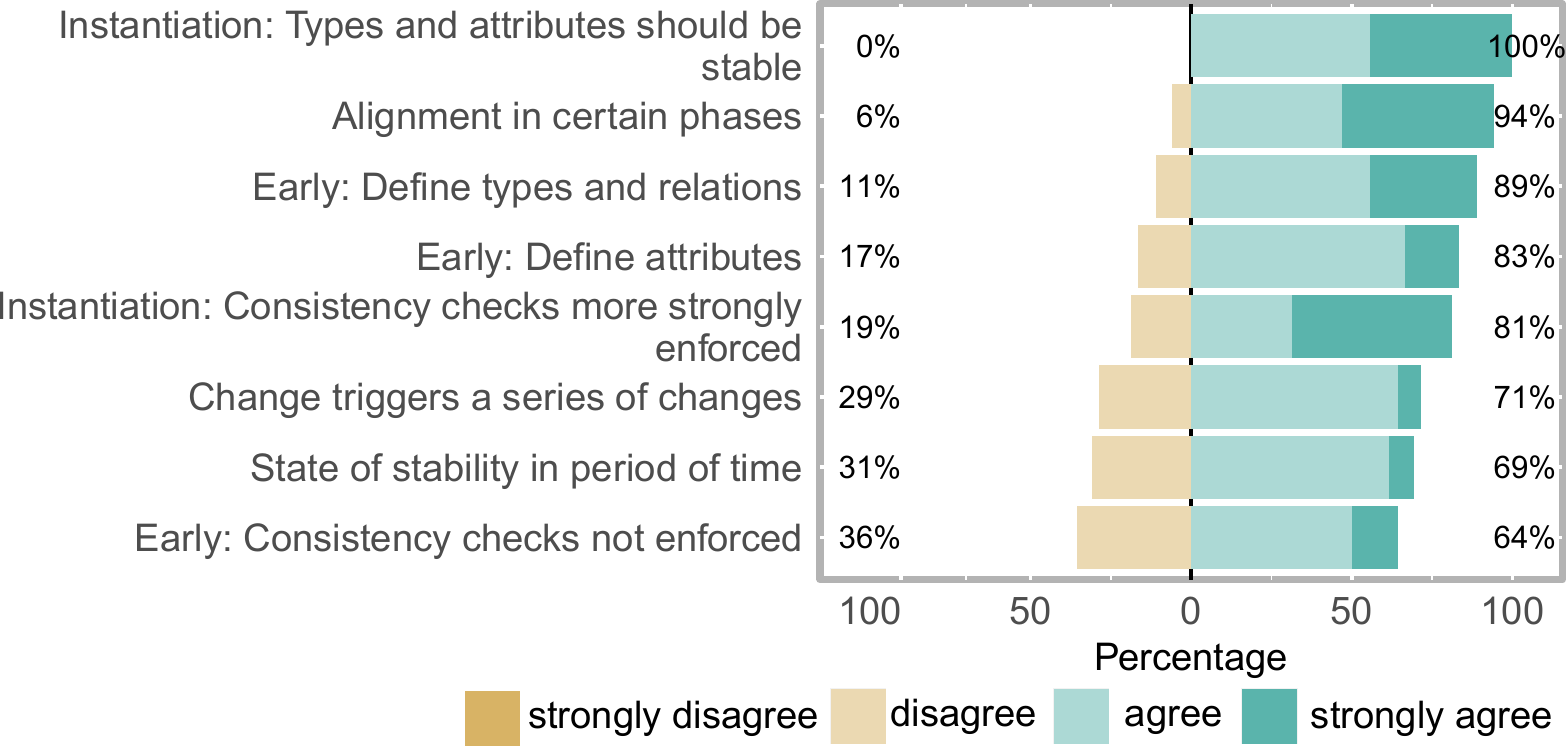}
	\caption{Survey responses regarding change of RIMs (n = 19)}
	\label{fig:changes}
\end{figure}

This section answers RQ3: \textit{\rqActions}
To answer this research question, we analyzed what actions our participants described in
the development lifecycle of their RIMs.
We found that they relate to different phases, from the RIMs' initial creation until the deprecation of elements.
We describe our findings based on interviews, survey responses, systems engineering tool data, and documentation.
Figure~\ref{fig:changes} shows the survey responses regarding RQ3.

\begin{center}
	\begin{small}
		\begin{tabular}{|p{.9\columnwidth}|}
			\hline
			When balancing alignment and diversity, we observe that practitioners carefully relate the lifecycle of the RIM and the lifecycle of concrete requirements instantiations. 
			The lifecycle of concrete requirements requires diversity in early phases, 
			but alignment especially as the product is released.
			Alignment can be ensured by consistency checks, whereas practitioners 
			support diversity by evolving the RIM based on needs observed with concrete 
			requirements.
			\\ \hline
		\end{tabular}
	\end{small}
\end{center}
\begin{figure} [h]
	\centering
	\includegraphics[width=0.85\linewidth]{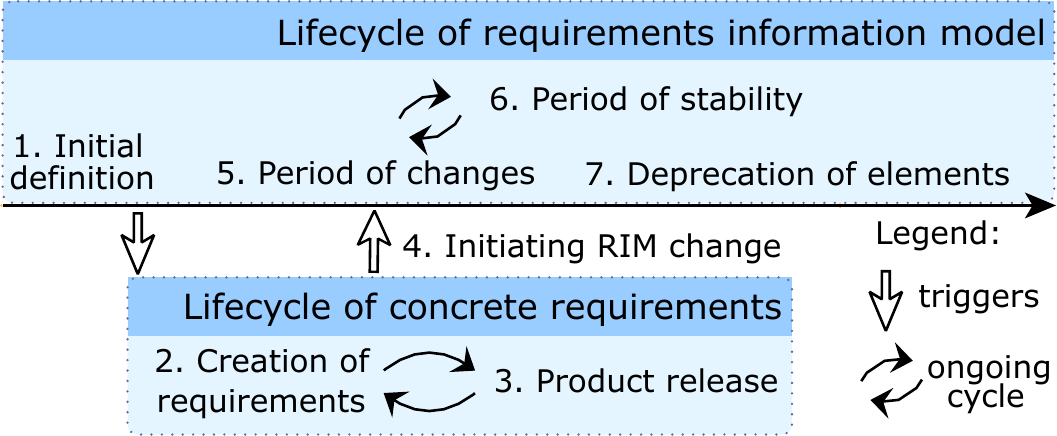}
	\caption{Lifecycles of RIMs and concrete requirements (letters refer to the subsections below)}
	\label{fig:actions}
\end{figure}

Figure~\ref{fig:actions} gives an overview of the lifecycles of RIMs and concrete requirements.
The RIM is initially defined and then instantiated for the creation of concrete requirements and product releases.
Based on the lifecycle of concrete requirements, periods of change and stability are triggered in the RIM, and eventually the deprecation of RIM elements.
Each phase in the figure is marked with the letter of the subsection describing it.
The following subsections will elaborate on the phases.
\subsection{Initial Definition of RIM}
The initial definition of the RIM sets up entity types and their relationships, as well as attributes.
For instance, at SUP, the early definition took between 6 to 12 months.
It was driven by tool experts within the company and support by a tool supplier.
All interviewees elaborated on their experiences with the initial definition of the RIM.
For tool suppliers, situations exist in which it is difficult to get good input for the early definition of the RIM:
\begin{aquote}{\daba}
	We come with proposals, don't get much feedback and then we launch the solutions and people start using it. But we don't get many definitions or needs or processes or anything.
\end{aquote}

\tefdde suggested to minimize the time between the early definition and the adoption by users.
According to this interviewee's experience, it is problematic if you ``want to measure and evaluate things exactly, rather than collecting feedback from users.''
Our respondents stated that early on entity types and relationships (89\%) and attributes (83\%) should be defined.
64\% of our participants stated that consistency checks are not enforced in this early stage.
\subsection{Creation of Concrete Requirements}
Once an initial definition of the RIM has been set up, users start adopting practices and instantiating it.
Three interviewees reported on the observed change as more and more users create requirements based on the RIM.
During the creation of concrete requirements, our companies use trainings to communicate new practices.
All of our survey respondents agree or strongly agree that entity types and attributes should be stable in this phase.
Consistency checks become more strongly enforced with time, as stated by 81\%.
However, as \ererwedf stated, in the beginning of the instantiation phase alignment is not strongly enforced and the goal is to be pragmatic:
\begin{aquote}{\ererwedf}
	We sometimes need to be quick and pragmatic and write requirements that are not so good right now, just to have something we can work with. And then we have to catch up in the end.
\end{aquote}

\subsection{Release of Concrete Requirements}
94\% of our participants agreed that alignment is more important in certain phases, e.g., when releasing the product.
Four interviewees explicitly stated that releases are planned as part of a start of production, but sometimes also as part of every sprint.
\begin{aquote}{\werer}
	If you have a very early prototype of something, you only want to get the principles right. If you are at the start of production, you need to have all details there. [...] More things should be enforced when you are closer to the start of production.
\end{aquote}
In the comments, an architect from an OEM pointed out that after a bigger release, when new development of a new platform is planned, the conceptual solution for requirements can be changed.
Otherwise, it should be stable.
\subsection{Initiating Change in a RIM} \label{sec:actions:initiatingChange}
With an overlap to the cycle in which a RIM is used to create concrete requirements, it also is changed by itself.
There are different ways to support change/refinement, as mentioned by 7 interviewees.
\tdfdd stressed that changes can happen due to the introduction of new 
technology, new standards, new methods, or a new organization.
An example of a change in the RIM of the systems engineering tool at OEM2 was when support for functional safety analysis was extended, which meant that new entity types (e.g., ``Hazardous 
Event'', and ``Safety Goal'') and their relationships were added to the RIM.
With changing technologies for vehicle messaging protocols (e.g., 
CAN, FlexRay, MOST), the metamodel is adjusted as well, for instance, to 
represent new or remove deprecated entity types.
A change with a greater impact was initiated as agile methods were introduced 
and organizations were restructured.
OEM2 created a dedicated team working with the RIM, analyzing the needs of 
different disciplines, and establishing a solution that also considered 
continuous integration.
In these situations, the ways of structuring functional requirements on a high 
level, keeping track of variability, and ensuring a traceable tool chain are 
revisited.
For instance, OEM2 changed the variability-related parts of the RIM.
Different ways of modeling are possible: Creating relationships from each 
requirement to the variants it is valid for, or managing variability models 
that link to the functions and requirements included in a certain variant.
Design decisions in the RIM are frequently rethought and can be implemented 
quickly, as the tool in use allows flexible changes to the RIM.

\trtretsewrt pointed out that the company tries to minimize change and avoid confusing the end users:
\begin{aquote}{\aLalal}
	The [RIM] was changed because new areas were added, for example, risk analysis. We have not changed the [RIM] all the time to not increase the confusion.
\end{aquote}

There are different approaches to initiating change.
At SUP, the tendency was rather to conduct ad-hoc changes and focused initiatives.
Committees have been used at OEM1 and OEM2, sometimes involving stakeholders from TOOL.

\subsubsection{Ad-hoc changes} Three interviewees gave examples of ad-hoc changes that they faced.
\tefdde from OEM2 stated that previously, metamodel changes were performed in an ad-hoc way, changes were done to see how they affected the usage, and then potentially reversed.
In small communities with 200--250 users, it is also easier to select whom to involve in decisions.
\begin{aquote}{\werer}
	If you have such a small community, you understand who is just picking on everything, and who is an expert in the subject.
\end{aquote}

\subsubsection{Committees} These groups evaluate a potential change and implement it when they are convinced of its quality.
Four interviewees reported on their experiences in committees.
\tdf mentioned that ``committees work more like a waterfall. It takes a long time to get decisions.''
At SUP, a community has been established to control change in a lightweight way: ``We have a community, put in the need to change [...], then people can comment and vote'' (\trtret).
\subsubsection{Focused Initiatives}
Initiatives are conducted for a limited period of time and focus on particular aspects of a RIM (e.g., variability).
Six interviewees had been actively involved in such initiatives.
\tefddedd mentioned that their initiative ``involved people from different departments'', but that it also has been ``a decision on management level'' to change the RIM in a certain way.
At OEM1, the involved participants worked with a test database to evaluate changes in a separate environment.
\tdf stated that key stakeholders are needed for successful initiatives: ``You need the `right' stakeholders, who understand alternative ways of working and can understand advantages and drawbacks of ideas.''
\subsection{Period of Change} \label{sec:actions:periodChange}
Periods of change happen as part of ad-hoc changes, work in committees, or focused initiatives.
Three interviewees explicitly mentioned that periods with series of changes exist.
This point was especially reported by interviewees from TOOL, having been involved in several endeavors to conduct change at several companies over the years.
71\% of our survey respondents stated that a change typically triggers a series of changes.
\tefdd explained how different people change the RIM in parallel, which is why they tried to modularize it.
\begin{aquote}{\ssdfwer}
	We know that our types and relationships will touch each other. We want to modularize the RIM. [...] And as long as we don't touch the interfaces, we can change things inside our modules.
\end{aquote}

Another task in periods of change is to refactor instantiated data.
\tdfdd stated that ``it is easy in the early phases of a project, but harder with more products.''
\subsection{Period of Stability} \label{sec:actions:periodStability}
69\% of the survey participants stated that a state of stability is reached after a series of changes.
\trtretsewrt stressed that ``there are different actors that perform changes until we come to a stable place. And that stable place might go through another iteration.''
The periods of stability are also related to the cycle of the instantiated data.
\trtretsewrt stated that ``after changing, usually the team that has requested the change is happy and reaches some stability in their work'', but that new changes arise to improve the alignment with the rest of the organization.
\tdfdd stated that periods of change are followed by periods of stability.
Initially, ``people are free, don't think formally, they try things out. And later on there is a shift in mentalities. Then version management gets more important.''
This implies that the desired characteristics of the RIM differ, depending on the position of requirements in the requirements lifecycle.
\tefddeddd mentioned that currently, the company is in a rather stable phase.
\begin{aquote}{\trtret}
	The [RIM] should be stable, we can maybe update 1-2 things, but not everything at once. We have most of the things in place now. We are in the phase that the changes are a handful, and the difficult thing is to make architects agree.
\end{aquote}

\subsection{Deprecation of Elements in RIM}
Three interviewees pointed to the need of deprecating unused elements in the RIM.
Rather than removing them, the companies prohibit the new creation of instances of the entity types:
\begin{aquote}{\ssdfwer}
	Deprecating things means that you cannot create them anymore. Old release data should be kept in the systems engineering tool, because sometimes you maybe have to touch it.
\end{aquote}
\trtretsewrt mentioned that one should understand how the RIM is instantiated and used, so that only unused elements are deprecated.
At SUP, for instance, 30 entity types are deprecated, whereas at OEM2, 27 deprecated types related to requirements exist.
Deprecated types often arise when stakeholders try out ways of modeling parts of the RIM and see the need to adjust it, while keeping the already instantiated data for maintenance purposes.

\subsection{Balancing Alignment and Diversity with RIMs}
The survey responses in Figure~\ref{fig:changes} indicate that our participants agree or strongly agree that types and attributes should be stable when the RIM is instantiated and concrete requirements are created (100\%) and that consistency checks should be more strongly enforced at that point in time (81\%).
Also the fact that alignment is more important in certain phases was agreed with (94\%).
We found a discrepancy in the answers related to consistency checks not being enforced in early phases: 83\% of the tools and methods experts agreed or strongly agreed with the statement, while 25\% of those who are no tools and methods expert agreed or strongly agreed with it.
Those who are not tools and methods experts are not involved in the early definition of the RIM and therefore potentially interpret ``early phases'' differently.

% !TEX root =  main.tex
\section{RQ4: Suggestions for Managing RIMs}  \label{sec:suggestions}
\begin{figure}
	\centering
	\includegraphics[width=\linewidth]{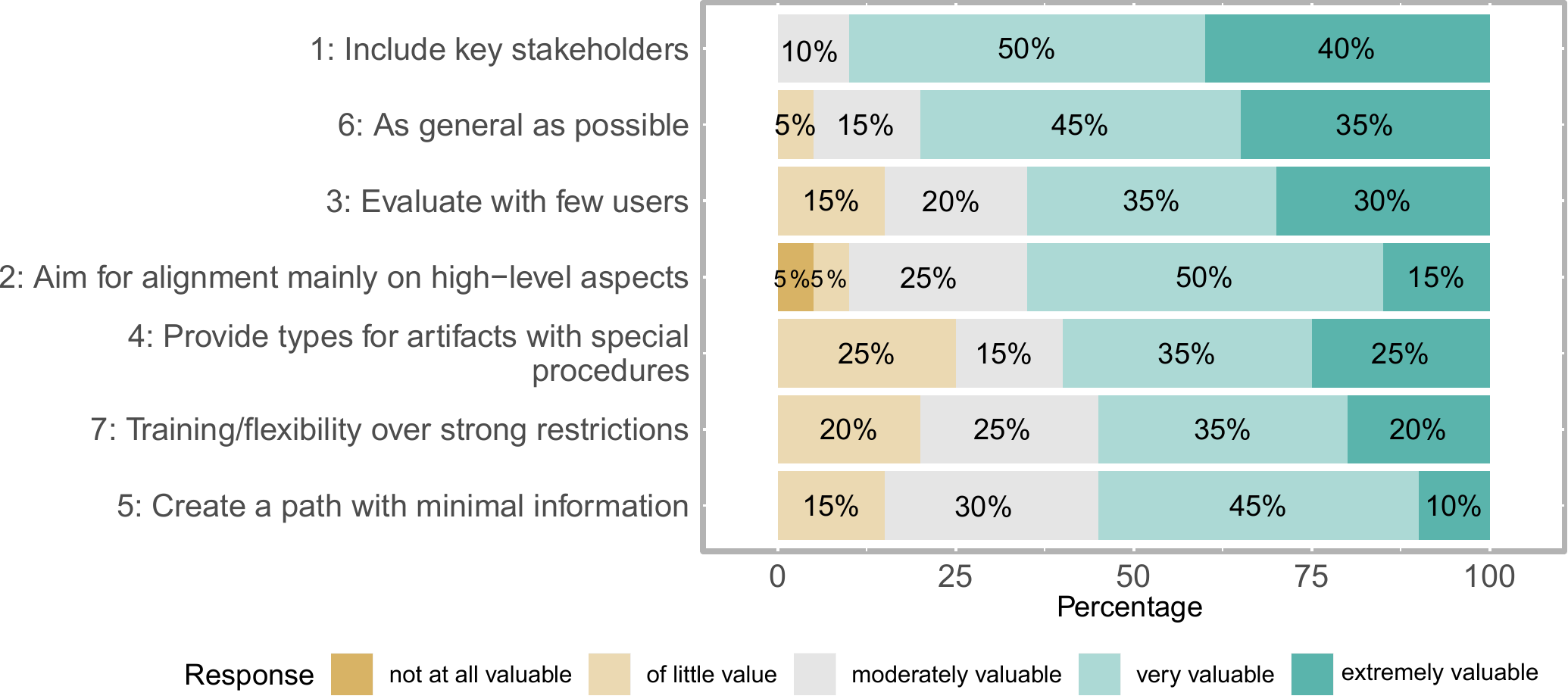}
	\caption{Survey responses regarding suggestions to manage RIMs (n = 19)}
	\label{fig:guidelines}
\end{figure}

This sections answers RQ4: \textit{\rqSuggestions}
Figure~\ref{fig:guidelines} shows an overview of the suggestions with the participants' ranking.
In this section, we report on the suggestions based on our data from interviews and the survey.
\subsection{Include Key Stakeholders}
One of the reported concerns when initiating change is to find suitable participants.
Six participants stated that it is good to involve end users, especially when changing the RIM.
Finding people who are not ``just picking on everything'' (\werer from OEM2), but understand trade-offs, is important for a successful balance of alignment and diversity.
We capture this advice in the first suggestion:
\begin{suggestion}
	Make sure you know the key stakeholders and include them to understand early what parts need to be aligned.
\end{suggestion}
89\% of our survey respondents ranked this suggestion as very or extremely valuable.
In the comments, a tools and methods expert stated that ``with more than 3 stakeholders, early development slows down.''
\subsection{Aim for Alignment Mainly on High-Level Aspects}
A RIM can include entity types on different levels of abstraction.
For instance, Functional Requirements are on a higher level than Software Requirements.
An interviewee argued that ``it is important to keep alignment on the top levels but allow variability on the lower levels'' (\ererwe).
We reflect this in the following suggestion:
\begin{suggestion}
	Establish a common (aligned) structure to organize high-level functionality and requirements of your product, but allow different modeling styles on a lower level.
\end{suggestion}
67\% of our survey respondents ranked this suggestion as very or extremely valuable.
An expert from a tooling company ranked the suggestion as not at all valuable and stated that ``for the lower levels [alignment is] even more important since it is required to be able to generate machine readable output.''
We see the benefit in using information to generate other artifacts and the need to have this prescriptive information in a suitable, aligned form.
However, for the aligned understanding of RIMs across team borders as a boundary object, high-level information was reported to be more relevant.

\subsection{Evaluate RIM Changes with Few Users}
During the initial definition of the RIM, stakeholders are more flexible than when data has already been instantiated based on the RIM.
Our interviewees stated that periods of stability make change more difficult (Section~\ref{sec:actions:periodStability}).
For this reason, three interviewees suggested to evaluate changes with a few users to avoid unnecessary changes in the future.
\begin{suggestion}
	Evaluate changes in the RIM in a small group of users, because changes will become more difficult with time.
\end{suggestion}
72\% of our respondents ranked this suggestion as very or extremely valuable.
In the comments, it was stressed that the group of involved users should be diverse.
\subsection{Provide Entity Types for Artifacts with Special Procedures}
When discussing the need for alignment based on standards (see Section~\ref{sec:why:alignment:standards}), our interviewees stressed that information with special procedures should be specifically classified in the RIM.
For instance, artifacts produced during safety analyses are created and reviewed with particular processes and should be easily identifiable in the tool.
Also, the classification of non-functional requirements can be used to derive the need for additional tests.

\begin{suggestion}
	Consider creating a separate requirement type or attribute for requirements that need special testing\slash safety\slash release procedures.
\end{suggestion}
This suggestion was ranked as very or extremely valuable by 61\% of the respondents.
In the comments, examples were given for procedures depending on whether a requirement is a safety requirement or not.
Of the respondents employed at a supplier, 17\% ranked this guideline as very or extremely valuable, whereas 67\% of the OEM employees ranked the guideline as very or extremely valuable.
The created procedures and methods within the companies differ and OEMs have a wider spectrum of methods, e.g., for integration tests on several levels.
\subsection{Create a Path with Minimal Information}
\trtretsewrt stated that depending on the complexity of a function, more or less information is needed, e.g., with respect to the detail of the specification of alternative scenarios.
\tefdde phrased this point as ``a short path'' to make people fill in relevant information in an easy way.

The interviewee gave an example of interfaces to other teams' artifacts that need to be defined as part of the short path.
However, other relationships might be optional and not used by every team in an organization.
\begin{suggestion}
	In the RIM, define a minimal amount of information that needs to be filled in (e.g., attributes of requirements), but allow users to add more details later if needed.
\end{suggestion}

This suggestion was found to be at least moderately valuable by 83\% of the respondents.
In the comments, two participants warned that if the minimal amount of information is too limited, too large differences between practices could arise.
\subsection{Aim for High Genericity}
We found that in some cases as a consequence of changes, new subtypes are created that have the same relationships and attributes as the super-types.
Three interviewees regarded that as a suboptimal solution.
\tefddeddd suggested ``to only create subtypes if you have different attributes, otherwise use the higher-level type.''
\begin{suggestion}
	%Try to k
	Keep the RIM as general as possible. Create subtypes %in the RIM 
	only if they possess special attributes or relationships.
\end{suggestion}
This suggestion was considered very or extremely valuable by 78\% of the respondents.
A tools and methods expert stated that this suggestion is risky if an entity type is used ``for various purposes.''
\subsection{Favor Training and Flexibility over Strong Restrictions}
Generic relationships are problematic when they are used extensively and it would be wiser to use typed relationships instead.
At SUP, one idea was to prohibit generic relationships.
\tefddeddd ``cannot see how to force people to do it right'' by disabling tool features.
The interviewee saw the need to ``have a better discussion of what is good.''
Four interviewees suggested to focus on training and communication.

\begin{suggestion}
	Align practices via training and communication instead of restricting the RIM too strongly.
\end{suggestion}
56\% of our survey respondents ranked this suggestion as very or extremely valuable.
Of those who were tools and methods experts, 62\% considered this guideline very or extremely valuable.
On the other hand, of those who were no tools and methods experts, 43\% considered this guideline very or extremely valuable.
A developer stressed that tool users should be educated as early as possible.
% !TEX root =  main.tex
\section{Discussion}  \label{sec:Discussion}
This paper extends the body of knowledge on how requirements information models are evolved in practice to balance alignment and diversity of requirements engineering practices in automotive companies.

\subsection{Reasons for Alignment and Diversity}
We explored reasons to support 
alignment and diversity of RE practices to answer \textit{RQ1} (Section~\ref{sec:why}):

\textbf{Summary (RQ1):} Alignment is mostly motivated by the need to facilitate integration, establish a common language, increase the quality of requirements, and adhere to standards. 
Diversity is mostly motivated by the variety of disciplines involved in automotive engineering, the methods, natures of functions, and different techniques for elicitation.

There exist only few approaches in the related literature that consider and support diverse or tailored practices in the context of RIMs (e.g.,~\citet{Weber2002}).
Generally, the assumption is that one common RIM can be created and used within an organization---and our findings confirm that there are indeed good reasons for it.
While our survey respondents generally agreed with the reasons for alignment, there were more mixed answers on reasons for diversity.
We found differences in the survey responses of OEM employees and respondents working at a supplier: For instance, different development methods were considered a reason for diversity by more OEM employees, while supplier employees more commonly regarded functions with different characteristics as a reason.
Clearly, the context and the organizational setting influence what the exact underlying reasons for diversity and alignment are.

\subsection{Enablers of Alignment and Diversity}
To our knowledge, consolidation of these concerns in a RIM has not received attention of research so far (our \emph{RQ2}).
We found that RIMs can enable the balance of alignment and diversity of RE practices (Section~\ref{sec:enablers}).

\textbf{Summary (RQ2):} RIMs support alignment by allowing to specify entity types and relationships, establishing common attributes, consistency checks, maturity levels, and Definition of Done criteria.
Diversity is enabled by supporting generic relationships, creating new subtypes with time, providing free text fields, and supporting several ways of organizing backlogs and projects.
While many of these enabling mechanisms were known before, our study sheds light on how they can be leveraged for the purpose of balancing alignment and diversity.
Related work on information models proclaims the need to capture concerns of various stakeholders using specialized entity types and relationships (e.g.,~\cite{Braun2005}).
The lack of DoD criteria was found to be problematic and compromising a shared understanding, which stresses their importance for alignment~\cite{Moe2012}.
Moreover, related work confirms the need to create traceability for artifacts of diverse types with meaningful link types~\cite{Gotel1994}.

\subsection{Actions to Balance Alignment and Diversity}
Even though this trade-off is difficult to manage, we found a set of actions \emph{(RQ3)} for systematic balancing between the two extremes.

\textbf{Summary (RQ3):} 
The lifecycle of concrete requirements influences the lifecycle of RIMs and how they are changed (Section~\ref{sec:actions}).
Existing requirements engineering approaches are (at least implicitly)  based on the assumption that their forms do not change.
Our findings examine how RIMs are created, extended, and evolved over time at three companies in automotive, supported by a tool supplier.
Concrete actions include the initial definition of the RIM during several months, ad-hoc changes, changes in committees or in focused initiatives, releases, and the deprecation of elements in the RIM.

Our findings suggest that, in practice, alignment is actually more enforced at later stages of the requirements lifecycle, when all requirements should be of consistently high quality.
After an initial definition of the RIM, as requirements are created and  products released, the RIM undergoes periods of change and stability, and elements are potentially deprecated.
The phases relate to some of the process activities described by John et al.~\cite{John1999}, but are not only based on concrete requirements, but also on how the RIM evolves.
While several RIMs have been proposed by related work, there is a lack of focus on how RIMs are changed and refined throughout their lifecycles.
To support agile methods and organizational change in practice, the need to evolve tool support and processes has been identified~\cite{Shahrokni2016}, in particular, when adopting model-driven engineering~\cite{Hutchinson2011}.
When evolving RIMs and managing change, \textit{model merging} can support the alignment of models created and changed by distributed teams~\cite{Brunet2006}.
In the future, longitudinal studies can be conducted to investigate in-depth what actions practitioners take and how they manifest in RIMs and systems engineering tool data.
\subsection{Suggestions for Balancing Alignment and Diversity}
\textbf{Summary (RQ4):} Our suggestions (Section~\ref{sec:suggestions}) are to include key stakeholders, evaluate changes with few users, and focus on the alignment of high-level aspects.
The importance of connecting requirements to the product level has been raised before~\cite{Gorschek2006}.
New entity types should only be created with good reasons (e.g., if special procedures, attributes, or relationships exist) and training and flexibility appear more beneficial than strong restrictions.
Moreover, we suggest to create a path with minimal information---allowing stakeholders to establish the core requirements early on and to extend them when more knowledge has been gathered.
This suggestion relates to Waterman's suggestions of keeping designs simple and delaying decision making, but planning for options in the area of agile architecture~\cite{Waterman2018}.
Such a path of minimal information in the RIM can also support agile development in large-scale automotive companies.

\subsection{Impact on Practice and Research}
\textit{Impact for practitioners:} The provided insights from four companies show how mechanisms in RIMs can help to address practical needs, what underlying reasons for alignment and diversity need to be balanced, and how diversity and alignment can be enabled by RIMs.
Our study helps stakeholders to see RIMs not as a rigid structure, but understand their RIMs' lifecycles and what actions can be taken to achieve a balance between diversity and alignment.
Moreover, practitioners can leverage the suggestions and use them to manage the balance of alignment and diversity in their organizations.

\textit{Impact for researchers:} Our study provides a better understanding of the practical trade-off of alignment and diversity.
We contributed to the knowledge base by investigating the evolution of RIMs over time and how they can be used to support diverse and aligned RE practices.
The concrete motivations, practices, and causalities raised here can facilitate future research.
As agile methods with their focus on reflection and continuous improvement become more common, also the need to evolve tool support and information models arises.
We hope to inspire research on creating methods and techniques to support the evolution and analyze the instantiation of RIMs.

While our findings are based on data that we collected within the automotive domain, we expect several of the findings to also be transferable to other large-scale systems and software engineering contexts.
Due to the large variety of disciplines in automotive~\cite{Weber2002,Ebert2017,Broy2007}, the heterogeneity of functions, and the supplier-OEM relationships, the need for diversity appears to be even more pronounced than in other industries.
Future studies will examine the applicability of our findings and suggestions in other domains.
\section{Conclusions and Outlook}
As organizations scale up and multiple teams conduct software and systems engineering in distributed setups, alignment and diversity of RE practices becomes an important topic.
The trade-off of alignment and diversity is directly observable in requirements information models (RIMs), as they manifest the common or diverse view of requirements and serve as boundary objects.
This paper explored the phenomenon of alignment and diversity in RIMs, including underlying reasons, enabling factors, actions that practitioners take, and suggestions for managing RIMs to balance alignment and diversity in large-scale automotive contexts.
A key observation relates to the role of the lifecycle of the requirements information model, and of the concrete requirements (instantiating concepts of the RIM). 
A suitable RIM should not overspecify and limit RE practices where it is not necessary.
The necessity for diversity appears to be strongest early in the requirements 
lifecycle, while the necessity for alignment becomes strongest close to the 
release.
Moreover, practitioners struggle with balancing need for stability of the metamodel to enable RE practices and the need to keep the RIM up to date with changing needs.
With a slow release cycle, periods of stability and change can be aligned with the concrete requirements lifecycle.
We foresee a future with more rapid release cycles that will also have stronger demands on the evolution of the RIM.
Our findings indicate that such a future would benefit from better support for such evolution.
\section*{Acknowledgments}
We are very grateful for the support of the participants involved in this study.

This work was partially supported by the Software Center Project 27 on RE for Large-Scale Agile System Development and by the Wallenberg AI, Autonomous Systems and Software Program (WASP) funded by the Knut and Alice Wallenberg Foundation.


\begin{thebibliography}{59}
	\expandafter\ifx\csname natexlab\endcsname\relax\def\natexlab#1{#1}\fi
	\providecommand{\url}[1]{\texttt{#1}}
	\providecommand{\href}[2]{#2}
	\providecommand{\path}[1]{#1}
	\providecommand{\DOIprefix}{doi:}
	\providecommand{\ArXivprefix}{arXiv:}
	\providecommand{\URLprefix}{URL: }
	\providecommand{\Pubmedprefix}{pmid:}
	\providecommand{\doi}[1]{\href{http://dx.doi.org/#1}{\path{#1}}}
	\providecommand{\Pubmed}[1]{\href{pmid:#1}{\path{#1}}}
	\providecommand{\bibinfo}[2]{#2}
	\ifx\xfnm\relax \def\xfnm[#1]{\unskip,\space#1}\fi
	%Type = Article
	\bibitem[{Bhat et~al.(2006)Bhat, Gupta and Murthy}]{Bhat2006}
	\bibinfo{author}{Bhat, J.M.}, \bibinfo{author}{Gupta, M.},
	\bibinfo{author}{Murthy, S.N.}, \bibinfo{year}{2006}.
	\newblock \bibinfo{title}{Overcoming requirements engineering challenges:
		Lessons from offshore outsourcing}.
	\newblock \bibinfo{journal}{IEEE Software} \bibinfo{volume}{23},
	\bibinfo{pages}{38--44}.
	\newblock \DOIprefix\doi{10.1109/MS.2006.137}.
	%Type = Inproceedings
	\bibitem[{Boulanger and Dao(2008)}]{Boulanger2008}
	\bibinfo{author}{Boulanger, J.L.}, \bibinfo{author}{Dao, V.Q.},
	\bibinfo{year}{2008}.
	\newblock \bibinfo{title}{Requirements engineering in a model-based methodology
		for embedded automotive software}, in: \bibinfo{booktitle}{2008 IEEE
		International Conference on Research, Innovation and Vision for the Future in
		Computing and Communication Technologies}, \bibinfo{publisher}{IEEE}. pp.
	\bibinfo{pages}{263--268}.
	\newblock \DOIprefix\doi{10.1109/RIVF.2008.4586365}.
	%Type = Book
	\bibitem[{Bowker and Star(1999)}]{Bowker1999}
	\bibinfo{author}{Bowker, G.C.}, \bibinfo{author}{Star, S.L.},
	\bibinfo{year}{1999}.
	\newblock \bibinfo{title}{Sorting things out: classification and its
		consequences}.
	\newblock \bibinfo{publisher}{MIT Press}, \bibinfo{address}{Cambridge, Mass.}
	%Type = Inproceedings
	\bibitem[{Braun and Winter(2005)}]{Braun2005}
	\bibinfo{author}{Braun, C.}, \bibinfo{author}{Winter, R.},
	\bibinfo{year}{2005}.
	\newblock \bibinfo{title}{A comprehensive enterprise architecture metamodel and
		its implementation using a metamodeling platform}, in:
	\bibinfo{editor}{Desel, J.}, \bibinfo{editor}{Frank, U.} (Eds.),
	\bibinfo{booktitle}{Enterprise Modelling and Information Systems
		Architectures}, \bibinfo{publisher}{Gesellschaft f{\"u}r Informatik}. pp.
	\bibinfo{pages}{64--79}.
	%Type = Article
	\bibitem[{Braun et~al.(2014)Braun, Broy, Houdek, Kirchmayr, M{\"u}ller,
		Penzenstadler, Pohl and Weyer}]{Braun2014}
	\bibinfo{author}{Braun, P.}, \bibinfo{author}{Broy, M.},
	\bibinfo{author}{Houdek, F.}, \bibinfo{author}{Kirchmayr, M.},
	\bibinfo{author}{M{\"u}ller, M.}, \bibinfo{author}{Penzenstadler, B.},
	\bibinfo{author}{Pohl, K.}, \bibinfo{author}{Weyer, T.},
	\bibinfo{year}{2014}.
	\newblock \bibinfo{title}{Guiding requirements engineering for
		software-intensive embedded systems in the automotive industry}.
	\newblock \bibinfo{journal}{Computer Science --- Research and Development}
	\bibinfo{volume}{29}, \bibinfo{pages}{21--43}.
	\newblock \DOIprefix\doi{10.1007/s00450-010-0136-y}.
	%Type = Article
	\bibitem[{Broy et~al.(2007)Broy, Kr{\"{u}}ger, Pretschner and
		Salzmann}]{Broy2007}
	\bibinfo{author}{Broy, M.}, \bibinfo{author}{Kr{\"{u}}ger, I.H.},
	\bibinfo{author}{Pretschner, A.}, \bibinfo{author}{Salzmann, C.},
	\bibinfo{year}{2007}.
	\newblock \bibinfo{title}{Engineering automotive software}.
	\newblock \bibinfo{journal}{Proceedings of the IEEE} \bibinfo{volume}{95},
	\bibinfo{pages}{356--373}.
	\newblock \DOIprefix\doi{10.1109/JPROC.2006.888386}.
	%Type = Inproceedings
	\bibitem[{Brunet et~al.(2006)Brunet, Chechik, Easterbrook, Nejati, Niu and
		Sabetzadeh}]{Brunet2006}
	\bibinfo{author}{Brunet, G.}, \bibinfo{author}{Chechik, M.},
	\bibinfo{author}{Easterbrook, S.}, \bibinfo{author}{Nejati, S.},
	\bibinfo{author}{Niu, N.}, \bibinfo{author}{Sabetzadeh, M.},
	\bibinfo{year}{2006}.
	\newblock \bibinfo{title}{A manifesto for model merging}, in:
	\bibinfo{booktitle}{Proceedings of the 2006 International Workshop on Global
		Integrated Model Management}, \bibinfo{publisher}{ACM}, \bibinfo{address}{New
		York, NY, USA}. pp. \bibinfo{pages}{5--12}.
	\newblock \URLprefix \url{http://doi.acm.org/10.1145/1138304.1138307},
	\DOIprefix\doi{10.1145/1138304.1138307}.
	%Type = Incollection
	\bibitem[{Cheng and Atlee(2009)}]{Cheng2009}
	\bibinfo{author}{Cheng, B.H.}, \bibinfo{author}{Atlee, J.M.},
	\bibinfo{year}{2009}.
	\newblock \bibinfo{title}{Current and future research directions in
		requirements engineering}, in: \bibinfo{booktitle}{Design Requirements
		Engineering: A Ten-Year Perspective}. \bibinfo{publisher}{Springer}, pp.
	\bibinfo{pages}{11--43}.
	%Type = Book
	\bibitem[{Creswell(2008)}]{Creswell2008}
	\bibinfo{author}{Creswell, J.W.}, \bibinfo{year}{2008}.
	\newblock \bibinfo{title}{Research Design: Qualitative, Quantitative, and Mixed
		Methods Approaches}.
	\newblock \bibinfo{edition}{3} ed., \bibinfo{publisher}{Sage Publications Ltd.}
	%Type = Book
	\bibitem[{Davis(2013)}]{Davis2013}
	\bibinfo{author}{Davis, A.}, \bibinfo{year}{2013}.
	\newblock \bibinfo{title}{Just enough requirements management: where software
		development meets marketing}.
	\newblock \bibinfo{publisher}{Addison-Wesley}.
	%Type = Article
	\bibitem[{Davis and Zowghi(2005)}]{Davis2005}
	\bibinfo{author}{Davis, A.M.}, \bibinfo{author}{Zowghi, D.},
	\bibinfo{year}{2005}.
	\newblock \bibinfo{title}{Good requirements practices are neither necessary nor
		sufficient}.
	\newblock \bibinfo{journal}{Requirements Engineering} \bibinfo{volume}{11},
	\bibinfo{pages}{1 -- 3}.
	%Type = Inproceedings
	\bibitem[{Dings{\o}yr et~al.(2014)Dings{\o}yr, F{\ae}gri and
		Itkonen}]{Dingsøyr2014a}
	\bibinfo{author}{Dings{\o}yr, T.}, \bibinfo{author}{F{\ae}gri, T.E.},
	\bibinfo{author}{Itkonen, J.}, \bibinfo{year}{2014}.
	\newblock \bibinfo{title}{What is large in large-scale? a taxonomy of scale for
		agile software development}, in: \bibinfo{editor}{Jedlitschka, A.},
	\bibinfo{editor}{Kuvaja, P.}, \bibinfo{editor}{Kuhrmann, M.},
	\bibinfo{editor}{M{\"a}nnist{\"o}, T.}, \bibinfo{editor}{M{\"u}nch, J.},
	\bibinfo{editor}{Raatikainen, M.} (Eds.), \bibinfo{booktitle}{Product-Focused
		Software Process Improvement}, \bibinfo{publisher}{Springer International
		Publishing}, \bibinfo{address}{Cham}. pp. \bibinfo{pages}{273--276}.
	%Type = Article
	\bibitem[{Doerr et~al.(2004)Doerr, Paech and Koehler}]{Doerr2004}
	\bibinfo{author}{Doerr, J.}, \bibinfo{author}{Paech, B.},
	\bibinfo{author}{Koehler, M.}, \bibinfo{year}{2004}.
	\newblock \bibinfo{title}{Requirements engineering process improvement based on
		an information model}.
	\newblock \bibinfo{journal}{Proceedings of the IEEE International Conference on
		Requirements Engineering} ,
	\bibinfo{pages}{70--79}\DOIprefix\doi{10.1109/ICRE.2004.1335665}.
	%Type = Article
	\bibitem[{Easterbrook et~al.(2008)Easterbrook, Singer, Storey and
		Damian}]{Easterbrook2008}
	\bibinfo{author}{Easterbrook, S.}, \bibinfo{author}{Singer, J.},
	\bibinfo{author}{Storey, M.A.}, \bibinfo{author}{Damian, D.},
	\bibinfo{year}{2008}.
	\newblock \bibinfo{title}{Selecting empirical methods for software engineering
		research}.
	\newblock \bibinfo{journal}{Guide to Advanced Empirical Software Engineering} ,
	\bibinfo{pages}{285--311}\DOIprefix\doi{10.1007/978-1-84800-044-5_11}.
	%Type = Article
	\bibitem[{Ebert and Favaro(2017)}]{Ebert2017}
	\bibinfo{author}{Ebert, C.}, \bibinfo{author}{Favaro, J.},
	\bibinfo{year}{2017}.
	\newblock \bibinfo{title}{Automotive software}.
	\newblock \bibinfo{journal}{IEEE Software} \bibinfo{volume}{34},
	\bibinfo{pages}{33--39}.
	\newblock \DOIprefix\doi{10.1109/MS.2017.82}.
	%Type = Article
	\bibitem[{Finkelstein et~al.(1992)Finkelstein, Kramer, Nuseibeh, Finkelstein
		and Goedicke}]{Finkelstein1992}
	\bibinfo{author}{Finkelstein, A.}, \bibinfo{author}{Kramer, J.},
	\bibinfo{author}{Nuseibeh, B.}, \bibinfo{author}{Finkelstein, L.},
	\bibinfo{author}{Goedicke, M.}, \bibinfo{year}{1992}.
	\newblock \bibinfo{title}{Viewpoints: A framework for integrating multiple
		perspectives in system development}.
	\newblock \bibinfo{journal}{International Journal of Software Engineering and
		Knowledge Engineering} \bibinfo{volume}{2}, \bibinfo{pages}{31--57}.
	%Type = Inproceedings
	\bibitem[{Fucci et~al.(2018)Fucci, Palomares, Franch, Costal, Raatikainen,
		Stettinger, Kurtanovic, Kojo, Koenig, Falkner, Schenner, Brasca,
		M\"{a}nnist\"{o}, Felfernig and Maalej}]{Fucci2018}
	\bibinfo{author}{Fucci, D.}, \bibinfo{author}{Palomares, C.},
	\bibinfo{author}{Franch, X.}, \bibinfo{author}{Costal, D.},
	\bibinfo{author}{Raatikainen, M.}, \bibinfo{author}{Stettinger, M.},
	\bibinfo{author}{Kurtanovic, Z.}, \bibinfo{author}{Kojo, T.},
	\bibinfo{author}{Koenig, L.}, \bibinfo{author}{Falkner, A.},
	\bibinfo{author}{Schenner, G.}, \bibinfo{author}{Brasca, F.},
	\bibinfo{author}{M\"{a}nnist\"{o}, T.}, \bibinfo{author}{Felfernig, A.},
	\bibinfo{author}{Maalej, W.}, \bibinfo{year}{2018}.
	\newblock \bibinfo{title}{Needs and challenges for a platform to support
		large-scale requirements engineering: A multiple-case study}, in:
	\bibinfo{booktitle}{Proceedings of the 12th ACM/IEEE International Symposium
		on Empirical Software Engineering and Measurement (ESEM '18)},
	\bibinfo{publisher}{ACM}, \bibinfo{address}{New York, NY, USA}. pp.
	\bibinfo{pages}{19:1--19:10}.
	\newblock \DOIprefix\doi{10.1145/3239235.3240498}.
	%Type = Article
	\bibitem[{Gorschek and Wohlin(2006)}]{Gorschek2006}
	\bibinfo{author}{Gorschek, T.}, \bibinfo{author}{Wohlin, C.},
	\bibinfo{year}{2006}.
	\newblock \bibinfo{title}{Requirements abstraction model}.
	\newblock \bibinfo{journal}{Requirements Engineering} \bibinfo{volume}{11},
	\bibinfo{pages}{79--101}.
	\newblock \DOIprefix\doi{10.1007/s00766-005-0020-7}.
	%Type = Inproceedings
	\bibitem[{Gotel and Finkelstein(1994)}]{Gotel1994}
	\bibinfo{author}{Gotel, O.}, \bibinfo{author}{Finkelstein, A.C.W.},
	\bibinfo{year}{1994}.
	\newblock \bibinfo{title}{An analysis of the requirements traceability
		problem}, in: \bibinfo{booktitle}{RE'94}, pp. \bibinfo{pages}{94--101}.
	\newblock \DOIprefix\doi{10.1109/ICRE.1994.292398}.
	%Type = Article
	\bibitem[{{Groen} et~al.(2017){Groen}, {Seyff}, {Ali}, {Dalpiaz}, {Doerr},
		{Guzman}, {Hosseini}, {Marco}, {Oriol}, {Perini} and {Stade}}]{Groen2017}
	\bibinfo{author}{{Groen}, E.C.}, \bibinfo{author}{{Seyff}, N.},
	\bibinfo{author}{{Ali}, R.}, \bibinfo{author}{{Dalpiaz}, F.},
	\bibinfo{author}{{Doerr}, J.}, \bibinfo{author}{{Guzman}, E.},
	\bibinfo{author}{{Hosseini}, M.}, \bibinfo{author}{{Marco}, J.},
	\bibinfo{author}{{Oriol}, M.}, \bibinfo{author}{{Perini}, A.},
	\bibinfo{author}{{Stade}, M.}, \bibinfo{year}{2017}.
	\newblock \bibinfo{title}{The crowd in requirements engineering: The landscape
		and challenges}.
	\newblock \bibinfo{journal}{IEEE Software} \bibinfo{volume}{34},
	\bibinfo{pages}{44--52}.
	\newblock \DOIprefix\doi{10.1109/MS.2017.33}.
	%Type = Article
	\bibitem[{Hertzum(2004)}]{Hertzum2004}
	\bibinfo{author}{Hertzum, M.}, \bibinfo{year}{2004}.
	\newblock \bibinfo{title}{Small-scale classification schemes: A field study of
		requirements engineering}.
	\newblock \bibinfo{journal}{Computer Supported Cooperative Work (CSCW)}
	\bibinfo{volume}{13}, \bibinfo{pages}{35--61}.
	%Type = Article
	\bibitem[{Hochm{\"{u}}ller(1997)}]{Hochmuller1997}
	\bibinfo{author}{Hochm{\"{u}}ller, E.}, \bibinfo{year}{1997}.
	\newblock \bibinfo{title}{Requirements classification as a first step to grasp
		quality requirements}.
	\newblock \bibinfo{journal}{REFSQ 1997} , \bibinfo{pages}{133--144}.
	%Type = Inproceedings
	\bibitem[{Hutchinson et~al.(2011)Hutchinson, Rouncefield and
		Whittle}]{Hutchinson2011}
	\bibinfo{author}{Hutchinson, J.}, \bibinfo{author}{Rouncefield, M.},
	\bibinfo{author}{Whittle, J.}, \bibinfo{year}{2011}.
	\newblock \bibinfo{title}{Model-driven engineering practices in industry}, in:
	\bibinfo{booktitle}{Proceedings of the 33rd International Conference on
		Software Engineering}, \bibinfo{organization}{ACM}. pp.
	\bibinfo{pages}{633--642}.
	%Type = Article
	\bibitem[{Ihantola and Kihn(2011)}]{Ihantola2011}
	\bibinfo{author}{Ihantola, E.M.}, \bibinfo{author}{Kihn, L.A.},
	\bibinfo{year}{2011}.
	\newblock \bibinfo{title}{Threats to validity and reliability in mixed methods
		accounting research}.
	\newblock \bibinfo{journal}{Qualitative Research in Accounting \& Management}
	\bibinfo{volume}{8}, \bibinfo{pages}{39--58}.
	%Type = Article
	\bibitem[{Inayat et~al.(2015)Inayat, Salim, Marczak, Daneva and
		Shamshirband}]{Inayat2015}
	\bibinfo{author}{Inayat, I.}, \bibinfo{author}{Salim, S.S.},
	\bibinfo{author}{Marczak, S.}, \bibinfo{author}{Daneva, M.},
	\bibinfo{author}{Shamshirband, S.}, \bibinfo{year}{2015}.
	\newblock \bibinfo{title}{A systematic literature review on agile requirements
		engineering practices and challenges}.
	\newblock \bibinfo{journal}{Computers in Human Behavior} \bibinfo{volume}{51},
	\bibinfo{pages}{915--929}.
	\newblock \DOIprefix\doi{10.1016/j.chb.2014.10.046}.
	%Type = Article
	\bibitem[{{International Organization for Standardization}(2011)}]{ISO26262}
	\bibinfo{author}{{International Organization for Standardization}},
	\bibinfo{year}{2011}.
	\newblock \bibinfo{title}{Road vehicles -- functional safety}.
	\newblock \bibinfo{journal}{ISO26262:2011} .
	%Type = Techreport
	\bibitem[{{ISO/IEC TR 11179-2:2019}(2019)}]{ISO11179}
	\bibinfo{author}{{ISO/IEC TR 11179-2:2019}}, \bibinfo{year}{2019}.
	\newblock \bibinfo{title}{{Information technology -- Metadata registries (MDR)
			-- Part 2: Classification}}.
	\newblock \bibinfo{type}{Technical Report}. {International Organization for
		Standardization}.
	%Type = Article
	\bibitem[{John et~al.(1999)John, Hoffmann, Weber, Nagel and Thomas}]{John1999}
	\bibinfo{author}{John, G.}, \bibinfo{author}{Hoffmann, M.},
	\bibinfo{author}{Weber, M.}, \bibinfo{author}{Nagel, M.},
	\bibinfo{author}{Thomas, C.}, \bibinfo{year}{1999}.
	\newblock \bibinfo{title}{Using a common information model as a methodological
		basis for a tool-supported requirements management process}.
	\newblock \bibinfo{journal}{INCOSE International Symposium}
	\bibinfo{volume}{9}, \bibinfo{pages}{1437--1441}.
	\newblock \DOIprefix\doi{10.1002/j.2334-5837.1999.tb00327.x}.
	%Type = Inproceedings
	\bibitem[{Kassab(2014)}]{Kassab2014}
	\bibinfo{author}{Kassab, M.}, \bibinfo{year}{2014}.
	\newblock \bibinfo{title}{An empirical study on the requirements engineering
		practices for agile software development}, in:
	\bibinfo{booktitle}{Proceedings of the 40th Euromicro Conference Series on
		Software Engineering and Advanced Applications (SEAA 2014)},
	\bibinfo{publisher}{IEEE}. pp. \bibinfo{pages}{254--261}.
	\newblock \DOIprefix\doi{10.1109/SEAA.2014.77}.
	%Type = Book
	\bibitem[{Laplante(2017)}]{Laplante2017}
	\bibinfo{author}{Laplante, P.A.}, \bibinfo{year}{2017}.
	\newblock \bibinfo{title}{Requirements engineering for software and systems}.
	\newblock \bibinfo{publisher}{Auerbach Publications}.
	%Type = Book
	\bibitem[{Leffingwell(2007)}]{Leffingwell2007}
	\bibinfo{author}{Leffingwell, D.}, \bibinfo{year}{2007}.
	\newblock \bibinfo{title}{Scaling Software Agility: Best Practices for Large
		Enterprises (The Agile Software Development Series)}.
	\newblock \bibinfo{publisher}{Addison-Wesley Professional}.
	%Type = Book
	\bibitem[{Leffingwell(2011)}]{Leffingwell2011}
	\bibinfo{author}{Leffingwell, D.}, \bibinfo{year}{2011}.
	\newblock \bibinfo{title}{Agile Software Requirements: Lean Requirements
		Practices for Teams, Programs, and the Enterprise}.
	\newblock Agile Software Development Series, \bibinfo{publisher}{Addison-Wesley
		Professional}.
	%Type = Article
	\bibitem[{Liebel et~al.(2018)Liebel, Tichy, Knauss, Ljungkrantz and
		Stieglbauer}]{Liebel2018}
	\bibinfo{author}{Liebel, G.}, \bibinfo{author}{Tichy, M.},
	\bibinfo{author}{Knauss, E.}, \bibinfo{author}{Ljungkrantz, O.},
	\bibinfo{author}{Stieglbauer, G.}, \bibinfo{year}{2018}.
	\newblock \bibinfo{title}{Organisation and communication problems in automotive
		requirements engineering}.
	\newblock \bibinfo{journal}{Requirements Engineering} \bibinfo{volume}{23},
	\bibinfo{pages}{145--167}.
	\newblock \DOIprefix\doi{10.1007/s00766-016-0261-7}.
	%Type = Article
	\bibitem[{Likert(1932)}]{Likert1932}
	\bibinfo{author}{Likert, R.}, \bibinfo{year}{1932}.
	\newblock \bibinfo{title}{A technique for the measurement of attitudes}.
	\newblock \bibinfo{journal}{Archives of Psychology} \bibinfo{volume}{22},
	\bibinfo{pages}{5--55}.
	%Type = Article
	\bibitem[{M{\'{e}}ndez~Fern{\'{a}}ndez
		et~al.(2011)M{\'{e}}ndez~Fern{\'{a}}ndez, Lochmann, Penzenstadler and
		Wagner}]{Mendez2011}
	\bibinfo{author}{M{\'{e}}ndez~Fern{\'{a}}ndez, D.}, \bibinfo{author}{Lochmann,
		K.}, \bibinfo{author}{Penzenstadler, B.}, \bibinfo{author}{Wagner, S.},
	\bibinfo{year}{2011}.
	\newblock \bibinfo{title}{A case study on the application of an artefact-based
		requirements engineering approach}.
	\newblock \bibinfo{journal}{15th Annual Conference on Evaluation {\&}
		Assessment in Software Engineering (EASE 2011)} ,
	\bibinfo{pages}{104--113}\DOIprefix\doi{10.1049/ic.2011.0013}.
	%Type = Article
	\bibitem[{{M{\'{e}}ndez Fern{\'{a}}ndez} et~al.(2010){M{\'{e}}ndez
			Fern{\'{a}}ndez}, Penzenstadler, Kuhrmann and Broy}]{Mendez2010}
	\bibinfo{author}{{M{\'{e}}ndez Fern{\'{a}}ndez}, D.},
	\bibinfo{author}{Penzenstadler, B.}, \bibinfo{author}{Kuhrmann, M.},
	\bibinfo{author}{Broy, M.}, \bibinfo{year}{2010}.
	\newblock \bibinfo{title}{A meta model for artefact-orientation: Fundamentals
		and lessons learned in requirements engineering}.
	\newblock \bibinfo{journal}{MODELS 2010} ,
	\bibinfo{pages}{183--197}\DOIprefix\doi{10.1007/978-3-642-16129-2_14}.
	%Type = Article
	\bibitem[{M{\'{e}}ndez~Fern{\'{a}}ndez and Wagner(2015)}]{Mendez2015}
	\bibinfo{author}{M{\'{e}}ndez~Fern{\'{a}}ndez, D.}, \bibinfo{author}{Wagner,
		S.}, \bibinfo{year}{2015}.
	\newblock \bibinfo{title}{Naming the pain in requirements engineering: A design
		for a global family of surveys and first results from {G}ermany}.
	\newblock \bibinfo{journal}{Information and Software Technology}
	\bibinfo{volume}{57}, \bibinfo{pages}{616 -- 643}.
	\newblock \DOIprefix\doi{https://doi.org/10.1016/j.infsof.2014.05.008}.
	%Type = Article
	\bibitem[{Moe et~al.(2012)Moe, Aurum and Dyb{\aa}}]{Moe2012}
	\bibinfo{author}{Moe, N.B.}, \bibinfo{author}{Aurum, A.},
	\bibinfo{author}{Dyb{\aa}, T.}, \bibinfo{year}{2012}.
	\newblock \bibinfo{title}{Challenges of shared decision-making: A multiple case
		study of agile software development}.
	\newblock \bibinfo{journal}{Information and Software Technology}
	\bibinfo{volume}{54}, \bibinfo{pages}{853--865}.
	%Type = Article
	\bibitem[{Niu et~al.(2018)Niu, Brinkkemper, Franch, Partanen and
		Savolainen}]{Niu2018}
	\bibinfo{author}{Niu, N.}, \bibinfo{author}{Brinkkemper, S.},
	\bibinfo{author}{Franch, X.}, \bibinfo{author}{Partanen, J.},
	\bibinfo{author}{Savolainen, J.}, \bibinfo{year}{2018}.
	\newblock \bibinfo{title}{Requirements engineering and continuous deployment}.
	\newblock \bibinfo{journal}{IEEE software} \bibinfo{volume}{35},
	\bibinfo{pages}{86--90}.
	%Type = Article
	\bibitem[{Pearson and Saeed(1997)}]{Pearson1997}
	\bibinfo{author}{Pearson, S.}, \bibinfo{author}{Saeed, A.},
	\bibinfo{year}{1997}.
	\newblock \bibinfo{title}{Information structures for traceability for
		dependable avionic systems}.
	\newblock \bibinfo{journal}{TECHNICAL REPORT SERIES-UNIVERSITY OF NEWCASTLE
		UPON TYNE COMPUTING SCIENCE} .
	%Type = Inproceedings
	\bibitem[{Pretschner et~al.(2007)Pretschner, Broy, Kr\"{u}ger and
		Stauner}]{Pretschner2007}
	\bibinfo{author}{Pretschner, A.}, \bibinfo{author}{Broy, M.},
	\bibinfo{author}{Kr\"{u}ger, I.H.}, \bibinfo{author}{Stauner, T.},
	\bibinfo{year}{2007}.
	\newblock \bibinfo{title}{Software engineering for automotive systems: A
		roadmap}, in: \bibinfo{booktitle}{Future of Software Engineering (FOSE '07)},
	pp. \bibinfo{pages}{55--71}.
	\newblock \DOIprefix\doi{10.1109/FOSE.2007.22}.
	%Type = Article
	\bibitem[{{QSR International Pty Ltd}(2019)}]{NVivo2019}
	\bibinfo{author}{{QSR International Pty Ltd}}, \bibinfo{year}{2019}.
	\newblock \bibinfo{title}{{NVivo 12 Pro}}.
	\newblock \bibinfo{journal}{\url{https://www.qsrinternational.com/nvivo}} .
	%Type = Inproceedings
	\bibitem[{Rempel et~al.(2013)Rempel, M{\"a}der and Kuschke}]{Rempel2013}
	\bibinfo{author}{Rempel, P.}, \bibinfo{author}{M{\"a}der, P.},
	\bibinfo{author}{Kuschke, T.}, \bibinfo{year}{2013}.
	\newblock \bibinfo{title}{An empirical study on project-specific traceability
		strategies}, in: \bibinfo{booktitle}{Proceedings of the 21st IEEE
		International Requirements Engineering Conference (RE'13)}, pp.
	\bibinfo{pages}{195--204}.
	\newblock \DOIprefix\doi{10.1109/RE.2013.6636719}.
	%Type = Incollection
	\bibitem[{Rodr{\'\i}guez et~al.(2009)Rodr{\'\i}guez, Mora, Martin, O'Connor and
		Alvarez}]{Rodriguez2009}
	\bibinfo{author}{Rodr{\'\i}guez, L.C.}, \bibinfo{author}{Mora, M.},
	\bibinfo{author}{Martin, M.V.}, \bibinfo{author}{O'Connor, R.},
	\bibinfo{author}{Alvarez, F.}, \bibinfo{year}{2009}.
	\newblock \bibinfo{title}{Process models of sdlcs: comparison and evolution},
	in: \bibinfo{booktitle}{Handbook of Research on Modern Systems Analysis and
		Design Technologies and Applications}. \bibinfo{publisher}{IGI Global}, pp.
	\bibinfo{pages}{76--89}.
	%Type = Article
	\bibitem[{Runeson and H{\"{o}}st(2009)}]{Runeson2009}
	\bibinfo{author}{Runeson, P.}, \bibinfo{author}{H{\"{o}}st, M.},
	\bibinfo{year}{2009}.
	\newblock \bibinfo{title}{Guidelines for conducting and reporting case study
		research in software engineering}.
	\newblock \bibinfo{journal}{Empirical Software Engineering} ,
	\bibinfo{pages}{131--164}\DOIprefix\doi{10.1007/s10664-008-9102-8}.
	%Type = Article
	\bibitem[{Sch{\"{o}}n et~al.(2017)Sch{\"{o}}n, Thomaschewski and
		Escalona}]{Schon2017}
	\bibinfo{author}{Sch{\"{o}}n, E.M.}, \bibinfo{author}{Thomaschewski, J.},
	\bibinfo{author}{Escalona, M.J.}, \bibinfo{year}{2017}.
	\newblock \bibinfo{title}{Agile requirements engineering: A systematic
		literature review}.
	\newblock \bibinfo{journal}{Computer Standards {\&} Interfaces}
	\bibinfo{volume}{49}, \bibinfo{pages}{79--91}.
	\newblock \DOIprefix\doi{https://doi.org/10.1016/j.csi.2016.08.011}.
	%Type = Inproceedings
	\bibitem[{Sedano et~al.(2019)Sedano, Ralph and Péraire}]{Sedano2019}
	\bibinfo{author}{Sedano, T.}, \bibinfo{author}{Ralph, P.},
	\bibinfo{author}{Péraire, C.}, \bibinfo{year}{2019}.
	\newblock \bibinfo{title}{The product backlog}, in:
	\bibinfo{booktitle}{Proceedings of the 41th International Conference on
		Software Engineering}, pp. \bibinfo{pages}{200--211}.
	\newblock \DOIprefix\doi{10.1109/ICSE.2019.00036}.
	%Type = Article
	\bibitem[{{Serna M.} et~al.(2017){Serna M.}, {Bachiller S.} and {Serna
			A.}}]{Serna2017}
	\bibinfo{author}{{Serna M.}, E.}, \bibinfo{author}{{Bachiller S.}, O.},
	\bibinfo{author}{{Serna A.}, A.}, \bibinfo{year}{2017}.
	\newblock \bibinfo{title}{Knowledge meaning and management in requirements
		engineering}.
	\newblock \bibinfo{journal}{International Journal of Information Management}
	\bibinfo{volume}{37}, \bibinfo{pages}{155 -- 161}.
	\newblock \DOIprefix\doi{https://doi.org/10.1016/j.ijinfomgt.2017.01.005}.
	%Type = Inproceedings
	\bibitem[{Shahrokni et~al.(2016)Shahrokni, S{\"{o}}derberg, Gergely,
		Pelliccione, S{\"{o}}derberg and Pelliccione}]{Shahrokni2016}
	\bibinfo{author}{Shahrokni, A.}, \bibinfo{author}{S{\"{o}}derberg, J.},
	\bibinfo{author}{Gergely, P.}, \bibinfo{author}{Pelliccione, P.},
	\bibinfo{author}{S{\"{o}}derberg, J.}, \bibinfo{author}{Pelliccione, P.},
	\bibinfo{year}{2016}.
	\newblock \bibinfo{title}{Organic evolution of development organizations - an
		experience report}, in: \bibinfo{booktitle}{SAE World Congress and Exhibition
		- Model-Based Controls and Software Development}, pp. \bibinfo{pages}{1--9}.
	%Type = Article
	\bibitem[{Sommerville and Sawyer(1997)}]{Sommerville1997}
	\bibinfo{author}{Sommerville, I.}, \bibinfo{author}{Sawyer, P.},
	\bibinfo{year}{1997}.
	\newblock \bibinfo{title}{Viewpoints: principles, problems and a practical
		approach to requirements engineering}.
	\newblock \bibinfo{journal}{Annals of Software Engineering}
	\bibinfo{volume}{3}, \bibinfo{pages}{101--130}.
	\newblock \DOIprefix\doi{10.1023/A:1018946223345}.
	%Type = Incollection
	\bibitem[{Star(1989)}]{Star1989a}
	\bibinfo{author}{Star, S.L.}, \bibinfo{year}{1989}.
	\newblock \bibinfo{title}{Chapter 2 - the structure of ill-structured
		solutions: Boundary objects and heterogeneous distributed problem solving},
	in: \bibinfo{editor}{Gasser, L.}, \bibinfo{editor}{Huhns, M.N.} (Eds.),
	\bibinfo{booktitle}{Distributed Artificial Intelligence}.
	\bibinfo{publisher}{Morgan Kaufmann}, \bibinfo{address}{San Francisco (CA)},
	pp. \bibinfo{pages}{37--54}.
	%Type = Article
	\bibitem[{Star and Griesemer(1989)}]{Star1989}
	\bibinfo{author}{Star, S.L.}, \bibinfo{author}{Griesemer, J.R.},
	\bibinfo{year}{1989}.
	\newblock \bibinfo{title}{Institutional ecology, `translations' and boundary
		objects: Amateurs and professionals in berkeley's museum of vertebrate
		zoology, 1907-39}.
	\newblock \bibinfo{journal}{Social Studies of Science} \bibinfo{volume}{19},
	\bibinfo{pages}{387--420}.
	\newblock \DOIprefix\doi{10.1177/030631289019003001}.
	%Type = Book
	\bibitem[{Tesch(1990)}]{Tesch1990}
	\bibinfo{author}{Tesch, R.}, \bibinfo{year}{1990}.
	\newblock \bibinfo{title}{Qualitative Research: Analysis Types and Software
		Tools}.
	\newblock \bibinfo{publisher}{Falmer Press}, \bibinfo{address}{London}.
	\newblock \URLprefix \url{https://books.google.se/books?id=pBqrngEACAAJ}.
	%Type = Article
	\bibitem[{{The R Foundation}(2019)}]{RProject2019}
	\bibinfo{author}{{The R Foundation}}, \bibinfo{year}{2019}.
	\newblock \bibinfo{title}{The {R} project for statistical computing}.
	\newblock \bibinfo{journal}{\url{https://www.r-project.org/}} .
	%Type = Article
	\bibitem[{Waterman(2018)}]{Waterman2018}
	\bibinfo{author}{Waterman, M.}, \bibinfo{year}{2018}.
	\newblock \bibinfo{title}{Agility, risk, and uncertainty, part 1: Designing an
		agile architecture}.
	\newblock \bibinfo{journal}{IEEE Software} \bibinfo{volume}{35},
	\bibinfo{pages}{99--101}.
	\newblock \DOIprefix\doi{10.1109/MS.2018.1661335}.
	%Type = Inproceedings
	\bibitem[{Weber and Weisbrod(2002)}]{Weber2002}
	\bibinfo{author}{Weber, M.}, \bibinfo{author}{Weisbrod, J.},
	\bibinfo{year}{2002}.
	\newblock \bibinfo{title}{Requirements engineering in automotive
		development---experiences and challenges}, in: \bibinfo{booktitle}{RE'02},
	pp. \bibinfo{pages}{331--340}.
	\newblock \DOIprefix\doi{10.1109/ICRE.2002.1048546}.
	%Type = Book
	\bibitem[{Wohlin et~al.(2012)Wohlin, Runeson, H{\"{o}}st, Ohlsson, Regnell and
		Wessl{\'{e}}n}]{Wohlin2012}
	\bibinfo{author}{Wohlin, C.}, \bibinfo{author}{Runeson, P.},
	\bibinfo{author}{H{\"{o}}st, M.}, \bibinfo{author}{Ohlsson, M.C.},
	\bibinfo{author}{Regnell, B.}, \bibinfo{author}{Wessl{\'{e}}n, A.},
	\bibinfo{year}{2012}.
	\newblock \bibinfo{title}{Experimentation in software engineering}. volume
	\bibinfo{volume}{9783642290}.
	\newblock \bibinfo{publisher}{Springer, Berlin, Heidelberg}.
	\newblock \DOIprefix\doi{10.1007/978-3-642-29044-2}.
	%Type = Inproceedings
	\bibitem[{Wohlrab et~al.(2018)Wohlrab, Pelliccione, Knauss and
		Gregory}]{Wohlrab2018REFSQ}
	\bibinfo{author}{Wohlrab, R.}, \bibinfo{author}{Pelliccione, P.},
	\bibinfo{author}{Knauss, E.}, \bibinfo{author}{Gregory, S.C.},
	\bibinfo{year}{2018}.
	\newblock \bibinfo{title}{The problem of consolidating {RE} practices at scale:
		An ethnographic study}, in: \bibinfo{booktitle}{Requirements Engineering:
		Foundation for Software Quality}, \bibinfo{publisher}{Springer
		International}. pp. \bibinfo{pages}{155--170}.
	%Type = Article
	\bibitem[{Wohlrab et~al.(2019)Wohlrab, Pelliccione, Knauss and
		Larsson}]{Wohlrab2019JSME}
	\bibinfo{author}{Wohlrab, R.}, \bibinfo{author}{Pelliccione, P.},
	\bibinfo{author}{Knauss, E.}, \bibinfo{author}{Larsson, M.},
	\bibinfo{year}{2019}.
	\newblock \bibinfo{title}{Boundary objects and their use in agile systems
		engineering}.
	\newblock \bibinfo{journal}{Journal of Software: Evolution and Process}
	\bibinfo{volume}{31}, \bibinfo{pages}{e2166}.
	\newblock \URLprefix
	\url{https://onlinelibrary.wiley.com/doi/abs/10.1002/smr.2166},
	\DOIprefix\doi{10.1002/smr.2166},
	\href{http://arxiv.org/abs/https://onlinelibrary.wiley.com/doi/pdf/10.1002/smr.2166}{{\tt
			arXiv:https://onlinelibrary.wiley.com/doi/pdf/10.1002/smr.2166}}.
	
\end{thebibliography}
\end{document}